\documentclass[%
reprint,
superscriptaddress,
amsmath,
amssymb,aps,
prc
]{revtex4-2}

\usepackage{graphicx}
\usepackage{dcolumn}
\usepackage{bm}
\usepackage{enumitem}
\usepackage{xcolor}

\begin{document}

\newcommand{\hiro}[1]{\textcolor{green}{#1}}
\newcommand{\ig}[1]{\textcolor{red}{#1}}
\newcommand{\sg}[1]{\textcolor{blue}{#1}}

\preprint{APS/123-QED}

\title{Photoneutron cross section measurements on $^{208}$Pb \\ in the Giant Dipole Resonance region}

\author{I.~Gheorghe}\email{ioana.gheorghe@nipne.ro}
\affiliation{National Institute for Physics and Nuclear Engineering,
Horia Hulubei (IFIN-HH), 30 Reactorului, 077125 Bucharest-Magurele, Romania}

\author{S.~Goriely}
\affiliation{Institut d'Astronomie et d'Astrophysique, Universit\'e Libre de Bruxelles, Campus de la Plaine, CP-226, 1050 Brussels, Belgium}

\author{N.~Wagner}
\affiliation{Institut f\"ur  Kernphysik, Technische Universit\"at Darmstadt, Darmstadt, 64289, Germany}

\author{T.~Aumann}
\affiliation{Institut f\"ur  Kernphysik, Technische Universit\"at Darmstadt, Darmstadt, 64289, Germany}
\affiliation{GSI Helmholtzzentrum f\"ur Schwerionenforschung, 64291 Darmstadt,Germany}
\affiliation{Helmholtz Forschungsakademie Hessen f\"ur FAIR (HFHF), GSI Helmholtzzentrum f\"ur Schwerionenforschung, 64291 Darmstadt, Germany}

\author{M.~Baumann}
\affiliation{Institut f\"ur  Kernphysik, Technische Universit\"at Darmstadt, Darmstadt, 64289, Germany}

\author{P.~van Beek}
\affiliation{Institut f\"ur  Kernphysik, Technische Universit\"at Darmstadt, Darmstadt, 64289, Germany}

\author{P.~Kuchenbrod}
\affiliation{Institut f\"ur  Kernphysik, Technische Universit\"at Darmstadt, Darmstadt, 64289, Germany}

\author{H.~Scheit}
\affiliation{Institut f\"ur  Kernphysik, Technische Universit\"at Darmstadt, Darmstadt, 64289, Germany}

\author{D.~Symochko}
\affiliation{Institut f\"ur  Kernphysik, Technische Universit\"at Darmstadt, Darmstadt, 64289, Germany}

\author{T.~Ari-izumi}
\affiliation{Konan University, Department of Physics, 8-9-1 Okamoto, Higashinada, Kobe 658-8501,Japan}

\author{F.L.~Bello Garrote}
\affiliation{Department of Physics, University of Oslo, N-0316 Oslo, Norway}

\author{T.~Eriksen}
\affiliation{Department of Physics, University of Oslo, N-0316 Oslo, Norway}

\author{W.~Paulsen}
\affiliation{Department of Physics, University of Oslo, N-0316 Oslo, Norway}

\author{L.G.~Pedersen}
\affiliation{Department of Physics, University of Oslo, N-0316 Oslo, Norway}

\author{F.~Reaz}
\affiliation{Department of Physics, University of Oslo, N-0316 Oslo, Norway}

\author{V.~W.~Ingeberg}
\affiliation{Department of Physics, University of Oslo, N-0316 Oslo, Norway}

\author{S.~Belyshev}
\affiliation{Lomonosov Moscow State University, Faculty of Physics, 119991 Moscow, Russia}

\author{S.~Miyamoto}
\affiliation{Laboratory of Advanced Science and Technology for Industry, University of Hyogo, 3-1-2 Kouto, Kamigori, Ako-gun, Hyogo 678-1205, Japan}

\author{H. Utsunomiya} \email{hiro@konan-u.ac.jp}
\affiliation{Konan University, Department of Physics, 8-9-1 Okamoto, Higashinada, Kobe 658-8501, Japan}

\date{\today}

\begin{abstract}
Photoneutron reactions on $^{208}$Pb in the Giant Dipole Resonance energy region have been investigated at the $\gamma$-ray beam line of the NewSUBARU facility in Japan. The measurements made use of quasi-monochromatic laser Compton backscattering $\gamma$-ray beams in a broad energy range, from the neutron threshold up to 38 MeV, and of a flat-efficiency moderated $^3$He neutron detection system along with associated neutron-multiplicity sorting methods. We report absolute cross sections and mean photoneutron energies for the $^{208}$Pb$(\gamma,\,inX)$ reactions with $i$~=~1 to 4. The fine structure present in the $^{208}$Pb$(\gamma,\,n)$ cross sections at incident energies lower than 13~MeV has been observed. The photoabsorption cross section has been obtained as the sum of the $(\gamma,\,inX)$ reaction cross sections. By reproducing the measured ring-ratio values at excitation energies below the two neutron separation energy, we were able to extract estimations on the $^{208}$Pb$(\gamma,\,n)$ photoneutron energy spectra and on the partial photoneutron cross sections for leaving the residual $^{207}$Pb in its ground and first two excited states. The present results are compared with data from the literature and statistical model calculations.

\end{abstract}

\maketitle


\section{Introduction} \label{sec_intro}
 
Photonuclear data, describing the response of atomic nuclei to photons, find use in fundamental nuclear physics and in a wide range of applications~\cite{kawano_2020}. The photoabsorption cross section is used to directly determine the $\gamma$-ray strength function ($\gamma$SF)~\cite{goriely_2019}, the key ingredient for computing $\gamma$-ray cascades in nuclear reactions~\cite{koning_2023_talys,herman_2007_empire}. Nuclear reactions induced by high energy photons of 10--20~MeV are dominated by $E1$ excitations known as the Giant Dipole Resonance (GDR) and understood as collective oscillations of the protons against the neutrons~\cite{bracco_tamii_2019}. The photoabsorption cross section in the GDR energy range reveals information on important nuclear quantities~\cite{goriely_2020}, such as the symmetry energy, which is the restoring force against the separation of protons and neutrons.

The doubly magic $^{208}$Pb is a benchmark case for theoretical modeling of the electric dipole response in nuclei~\cite{goriely_2020}. It has been extensively investigated through photonuclear reactions in pioneering experiments at the Livermore~\cite{harvey_1964,berman_1987} and Saclay~\cite{veyssiere_1970} positron in flight annihilation facilities. Recently, fine structures superimposed on the broad GDR resonance have been observed in great detail in high energy resolution inelastic hadron scattering experiments at RCNP GrandRaiden~\cite{tamii_2011,poltoraska_2014} and iThemba~\cite{jingo_2018} facilities, revealing single particle manifestations. Based on the $(p,\,p')$ RCNP data and on $^{208}$Pb$(^3$He,~$^3$He$'\gamma)$ Oslo-type data, the total $^{208}$Pb $\gamma$SF and its $E1$, $M1$ and $E2$ components have been extracted in the 2.7~--~20~MeV excitation energy range~\cite{bassauer_2016}.

However, there are unresolved systematic discrepancies~\cite{varlamov_2021} between photoabsorption and photoneutron cross sections obtained at the Saclay and Livermore facilities, with Saclay~\cite{veyssiere_1970} $(\gamma,\,abs)$ and $(\gamma,\,n)$ cross sections systematically higher than the Livermore ones~\cite{harvey_1964}, and Livermore $(\gamma,\,2n)$ cross sections overestimating the Saclay ones. The different neutron multiplicity sorting procedures employed at the two laboratories have been suggested in Refs.~\cite{wolynec_1984,wolynec_1987} as a discrepancy source. The need to remeasure the $^{208}$Pb photoabsorption excitation function in the GDR energy range has also been pointed out in Ref.~\cite{goriely_2020} in order to shed light on a discontinuity in the $E1$ moments between neighboring $^{208}$Pb and $^{209}$Bi nuclei.

New measurements of photoneutron reactions on $^{208}$Pb in and above the GDR energy range have been performed at the laser Compton scattering (LCS) $\gamma$-ray beam line of the NewSUBARU synchrotron radiation facility in Japan~\cite{amano_2009,horikawa_2010}. Photoneutron cross sections have been measured at incident photon energies between 7.5 and 38 MeV. The investigations made use of a novel high-and-flat efficiency moderated neutron detection array (FED) and the associated neutron-multiplicity sorting methods~\cite{utsunomiya_2017,gheorghe_2021}.

Decay experiments which measure the energy and angular distribution of $\gamma$-rays and particles emitted from GDR states are in high demand for bringing information on the microscopic nature of the GDR~\cite{lv_niu_2021}. However, these are very scarce, because of low detection efficiencies of neutron time of flight experiments. In lack of direct spectroscopic measurements, the present experiment provides indirect determinations of the average energy of neutrons emitted in the photoneutron reactions on $^{208}$Pb. In the present work, we extract estimations for the neutron emission spectra in the low energy $^{208}$Pb$(\gamma,\,n)$ reaction. 

In Sect.~\ref{sec_exp_method}, we present the experimental technique and methodology, with focus on diagnostics of the incident LCS $\gamma$-ray beams and neutron detection. The data analysis methods concerning neutron multiplicity sorting, extraction of information on the photoneutrons mean energy and spectra, as well as the energy unfolding are discussed in Sect.~\ref{sec_data_analysis}. Results are discussed and compared with preceding data in Sect.~\ref{sec_results} and with theoretical calculations in Sect.~\ref{sec_STAT_calc}. 
These include the photoabsorption cross section, the photoneutron cross sections and average energies and also estimations on the $^{208}$Pb$(\gamma,\,n)$ photoneutron spectra and partial cross sections. A summary and conclusions are given in Sect.~\ref{sec_summary}.

\section{Experimental method} \label{sec_exp_method}

\begin{figure*}[t]
\centering
\includegraphics[width=0.9\textwidth, angle=0]{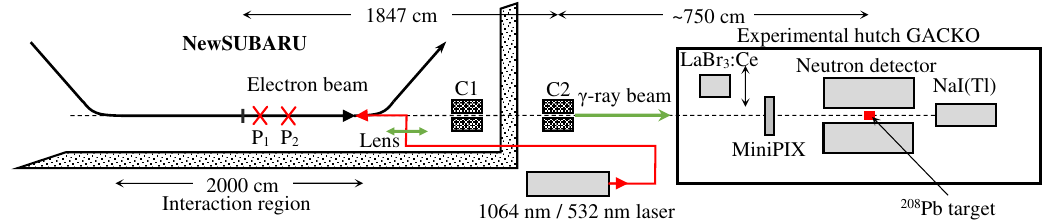}
\caption{Diagram showing the BL01 LCS $\gamma$-ray beam line and the experimental hutch GACKO at the NewSUBARU synchrotron radiation facility (not to scale). P$_1$ and the P$_2$ mark the focus point positions of the 1064~nm and, respectively the 532~nm laser beams at 1.8 and, respectively 3.8~m downstream of the eletron beam focus in the middle of the BL01.}\label{fig01_BL01_PRC_Pb208_24_crop}     
\end{figure*}

The apparatus used for the present experiment is shown schematically in Fig.~\ref{fig01_BL01_PRC_Pb208_24_crop}. Laser beams were guided into the BL01 straight beam-line section of the NewSUBARU synchrotron and were scattered by electrons circulating in the ring. The back-scattered $\gamma$-ray photons that passed through a system of two lead collimators (C$_1$ and C$_2$ in Fig.~\ref{fig01_BL01_PRC_Pb208_24_crop}) irradiated the target which was located in the GACKO (Gamma collaboration hutch of Konan University) experimental hutch, 7.5~m downstream of the second collimator. 

The energy resolution and flux of the $\gamma$-ray beam were monitored with a lanthanum bromide (LaBr$_3$:Ce) and a NaI:Tl detector, respectively. The reaction neutrons were moderated and recorded with an array of $^3$He counters placed in concentric rings around the target. The data were written in a triggerless list mode, using an eight-parameter 25 MHz digital data acquisition system which recorded the time and energy of the signals provided by the LaBr$_3$:Ce and NaI detectors, the neutron detection time in each ring of $^3$He counters and the external laser triggering signal. 

We note that the present experiment provides the $\sigma_{\gamma,\,inX}$ sum cross sections of all photoneutron reactions with $i$ neutrons in the final state, accompanied or not by charged particle emission:
\begin{align}
\sigma_{\gamma,\,inX} & \equiv \sigma(\gamma,\,inX) \nonumber \\
                      & = \sigma(\gamma,\,in) + \sigma(\gamma,\,inp) + \sigma(\gamma,\,in\alpha) + \dots       
\end{align}
However, the large Coulomb barrier in $^{208}$Pb~($\sim$26~MeV for $\alpha$ particles) hinders the emission of charged particles. Thus, for incident photon energies covering the GDR range up to $\sim$20~MeV, both the cross sections and the average photoneutron energies for the $(\gamma,\,inX)$ reactions generally coincide to or are a good approximation for the $(\gamma,\,in)$ neutron emission only ones~\cite{kawano_2020}. 

\subsection{LCS $\gamma$-ray beams}

Table~\ref{table_beams} gives the main properties of the electron, laser and $\gamma$-ray beams employed in the present study. The $\gamma$ beam energy was varied in 85 steps between 7.5 and 38.12 MeV by using two lasers with wavelengths of 1064 and 532 nm and tuning the electron energy between approximately 650 and 1050 MeV. The maximum energy of the LCS $\gamma$-ray beam is directly determined by the known laser wavelength and by the electron beam energy, which has been calibrated with an accuracy of (5.5~--~9.4)~$\times$~10$^{-5}$ \cite{utsunomiya_2014}. 

Interactions between the unsynchronized laser and electron beams could take place along the entire 20~m length of the electron beamline, according to the laser-electron beam spatial overlap (see Fig.~6 of Ref.~\cite{filipescu_2023}). The resulting LCS $\gamma$-ray beam had a pulsed time structure given by the $slow$ laser time structure (1~--~20~kHz, 20~--~40~ns pulse width) and the $fast$ electron beam one (500~MHz, 60~ps pulse width). A 100~ms macro-time structure of alternating 80~--~90 ms beam-on and 20~--~10 ms beam-off intervals was used for background monitoring, where Table~\ref{table_beams} gives the corresponding 80 or 90$\%$ fill factors for each measurement range. The NaI:Tl detector (8" diameter $\times$ 12" length) used as a flux monitor was placed in beam, downstream of the neutron detection system. The number of incident $\gamma$-rays was obtained by applying the pile-up or Poisson fitting method~\cite{utsunomiya_2018} on the NaI:Tl detector response functions, which represented summed spectra for the Poisson distributed $\gamma$ photons in the 20~--~40~ns wide pulses. The mean numbers of $\gamma$ rays per pulse and the incident photon flux values are given in Table~\ref{table_beams}. 

\begin{table}
\caption{\label{table_beams} Parameters for the laser, electron and LCS $\gamma$-ray beams and measurement conditions. $Z_R$ is the laser Rayleigh length~\cite{filipescu_2023}, $\Delta z$ is the longitudinal displacement between the focal positions of the laser and electron beams. $f_\mathrm{e^-}$ is the RF frequency of the NewSUBARU storage ring. We give the pulse width of the 198 electron beam bunches circulating in the ring. $E_{m}$ is the maximum energy of the $\gamma$-ray beam~\cite{utsunomiya_2014}.}
\begin{ruledtabular}
\begin{tabular}{lrr}
                  & $(\gamma,\,n)$    & $(\gamma,\,in)$  \\ 
 \multicolumn{2}{r}{neutron } &  multiplicity   \\                  
 \multicolumn{2}{r}{ counting} &   sorting  \\                  \hline 
 Laser beam:      &                   &                        \\                       
 Laser            & Inazuma           & Talon                  \\     
 Wavelength (nm)  & 1064              & 532                    \\  
 $Z_R$ (m)        & 0.57              & 6.1                    \\ 
 $\Delta z$ (m)   & 1.8               & 2.8                    \\ 
 Power (W)        & $<$40             & $<$20                  \\   
 $f_\mathrm{laser}$ (kHz)   & 20      & 1                      \\    
 Pulse width (ns)           & 20      & 40                     \\     
 Beam on/off fill factor (\%)           & 80      & 90                     \\ 
 \multicolumn{3}{c}{100\% linear polarization perpendicular to accelerator plane}    \\  \hline 
 Electron beam:             &         &                                              \\                        
 $E_\mathrm{e^-}$ (MeV)     & 651.30 -- 887.65     & 649.29 -- 1050.63               \\ 
 $f_\mathrm{e^-}$ \& pulse width        & \multicolumn{2}{c}{500 MHz \& 60 ps}       \\  
 \multicolumn{3}{l}{Emittance ($\varepsilon_x,\varepsilon_y$) (nm-rad): }   \\ 
 -- nominal at injection        & \multicolumn{2}{c}{(38,1-3.8)}       \\   
 -- simulated       & \multicolumn{2}{c}{(50,5) -- (70,7)}          \\ \hline
 LCS $\gamma$-ray beam:      &                   &                                   \\                        
 $E_{m}$ (MeV)    & 7.50 -- 13.86     & 13.84 -- 38.02                    \\ 
 Mean nb. of $\gamma$/pulse  & 5 -- 14           & 6 -- 20                           \\ 
 Incident flux ($\gamma$/s)  & (8 -- 22) $\times$ $10^4$     & (0.5 -- 1.8) $\times$ $10^4$    \\ 
 $\Delta$E$_\mathrm{FWHM}$ (MeV)  & 0.2 -- 0.5     & 0.2 -- 1.1                      \\ 
 $\Delta$E$_\mathrm{FWHM}$ (\%)   & 2.7 -- 3.9     & 1.6 -- 3.1                      \\ \hline
 \multicolumn{2}{l}{Measurement conditions:} &  \\
 Irradiation time (min) & 5~--~30 & $\sim$120 \\
 Target areal &   &   \\
 density (g/cm$^2$) & 4.36 & 10.97 \\ 
\end{tabular}
\end{ruledtabular}
\end{table}

\begin{figure}[t]
\centering
\includegraphics[width=0.98\columnwidth, angle=0]{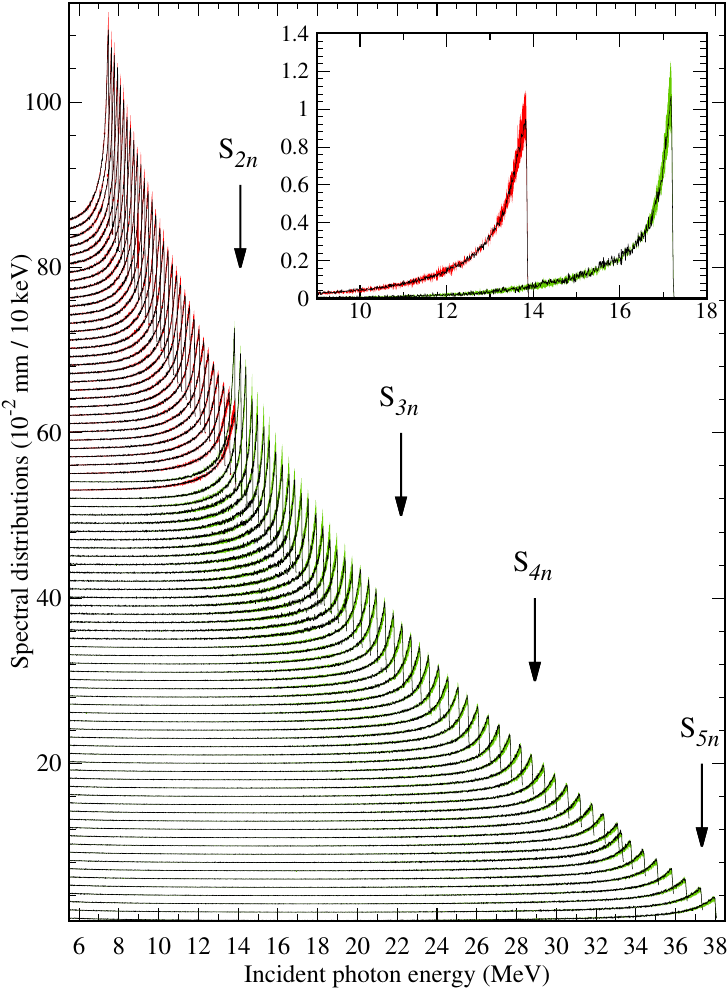} 
\caption{The simulated energy profiles of the 85 incident LCS $\gamma$-ray beams used in the present experiment. Each curve is offset along the vertical axis for clarity. The spectral uncertainty is shown by the red band for $\gamma$ beams obtained with the 1064~nm wavelength laser and by the green band for the ones obtained with the 532~nm laser. The inset shows the magnified spectra for the 13.86 and 17.21~MeV beams.}\label{fig_inc_spectra_lin2}     
\end{figure}

The spectral distribution of the incident LCS $\gamma$-ray beam was monitored in between target irradiations with a 3.5" diameter $\times$ 4.0" length LaBr$_3$:Ce detector placed in beam. For the incident spectra measurements, the laser was operated in a continuous wave (CW) mode at reduced power. The experimental detector response functions were reproduced by Monte Carlo simulations performed with the LCS $\gamma$-source simulation code $\texttt{eliLaBr}$~\cite{eliLaBr_github,filipescu_2023,filipescu_2022}. The simulated energy spectra of the incident beams used in the present experiment are shown in Fig.~\ref{fig_inc_spectra_lin2} by the $L(E_\gamma,E_m)$ path length weighted energy distributions for each LCS $\gamma$-ray beam of $E_m$ maximum energy~\cite{filipescu_2023}. 
The $L(E_\gamma,E_m)$ distributions account for the $\gamma$-beam self-attenuation in the irradiated sample material and for the secondary radiation generated by electromagnetic interaction of the $\gamma$-beam with the target, which, for $\sim$30--40~MeV photon beams, has sufficiently high energies to induce nuclear reactions in the target. 

By investigating the range of suitable values for the electron beam emittance and for small laser-electron beam transverse offsets, we estimated the uncertainty in the incident spectra determination, which is shown in Fig.~\ref{fig_inc_spectra_lin2} by the red and green bands for the $\gamma$ beams produced with the 1064 and respectively 532~nm lasers, respectively. Table~\ref{table_beams} gives values for the main input parameters used in the LCS $\gamma$-ray beam simulations and for the energy spread of the best fit $\gamma$-ray beams.

\subsection{Targets}

Enriched metal powder of $^{208}$Pb (98.4$\%$ $^{208}$Pb, 1.2$\%$ $^{207}$Pb, 0.1$\%$ $^{206}$Pb, 0.3$\%$ $^{204}$Pb) was pressed and shaped into two targets of 8~mm diameter and lengths of 4 and 6~mm. The neutron counting measurements below S$_{2n}$ were performed with the thinner target of 4.36~g/cm$^2$ areal density, for which the photon transmission varied between 79.4 and 82.1$\%$. The neutron multiplicity sorting measurements above S$_{2n}$ were performed with the two targets stacked, resulting in a 10.97~g/cm$^2$ total areal density. The photon transmission through the stacked targets varied between 45.0 and 55.3$\%$. The target alignment to the $\gamma$-beam axis has been performed using the visible synchrotron radiation passing through the collimators C1 and C2 and has been checked by measurements with an X-ray MiniPIX camera~\cite{minipix_website,granja_2022}, which also confirmed the beam spot estimations of $\sim$4~mm diameter~\cite{Ariizumi_2023}. 

\subsection{Neutron detection}

\begin{figure}[t]
\centering
\includegraphics[width=0.98\columnwidth, angle=0]{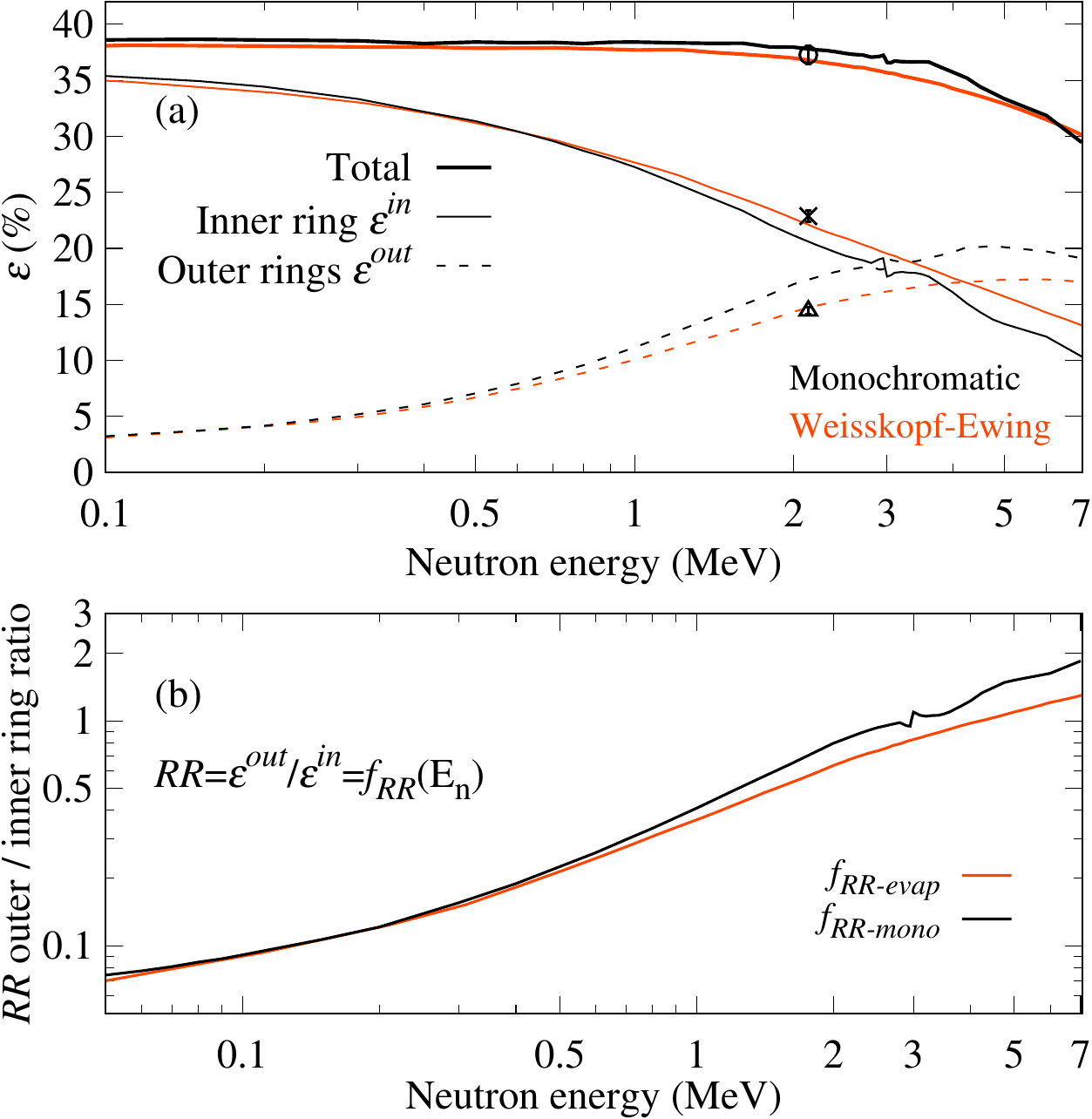} 
\caption{FED calibration. (a) Experimental $^{252}$Cf measurement for the total (circle), inner ring (cross) and summed outer rings (triangle) detection efficiencies reproduced by \textsc{mcnp} simulations for monochromatic neutrons (black) and e\-vap\-o\-ra\-tion neutron spectra (red). (b) The average neutron energy as function of the $\varepsilon^{in}/\varepsilon^{out}$ ratio between the efficiencies of the inner and summed outer rings of counters, here referred to as ring ratio functions $f_{RR}$ and computed for both monochromatic and e\-vap\-o\-ra\-tion neutron spectra.}\label{fig_fed_eff_new}     
\end{figure}

The $^{208}$Pb targets were placed in the center of a flat-efficiency moderated neutron detection array of $^3$He proportional counters (10 atm., 2.5 cm diameter $\times$ 45 cm length active volume)~\cite{utsunomiya_2017}. The array consisted of three concentric rings of 4, 9, and 18 $^3$He counters placed 5.5, 13.0 and 16.0 cm from the $\gamma$-ray beam axis, respectively (see Figs.~1 and 2 of Ref.~\cite{utsunomiya_2017}). The $^3$He tubes were embedded in a high-density polyethylene moderator block having a central axial hole that allowed the $\gamma$-ray beam to pass through.

Figure~\ref{fig_fed_eff_new}(a) shows \textsc{mcnp}~\cite{mcnp} simulations for the total detection efficiency (thick solid lines) and for the detection efficiencies of the inner ring (thin solid lines) and of the summed two outer rings (dotted lines) compared to an experimental calibration with a $^{252}$Cf source of known activity~\cite{utsunomiya_2017}. Calculations are shown for monochromatic neutrons (black) and for e\-vap\-o\-ra\-tion neutron spectra described by the Weisskopf-Ewing function~\cite{Weisskopf_1937} (red lines) and are represented at the corresponding average neutron energies. We notice the kink in the monochromatic efficiency curve given by the 3~MeV resonance in the cross section of neutron elastic scattering on $^{12}$C. The $^3$He array geometrical configuration was optimized so that the total detection efficiency is (i) insensitive to the shape of the neutron spectrum and (ii) constant in the 10~keV to 5~MeV average neutron energy range, with a $\sim$5$\%$ variation between 38$\%$ and 33$\%$ for both monochromatic and e\-vap\-o\-ra\-tion neutron spectra. 

However, the detection efficiencies of the inner and outer rings do depend on the neutron energy: the inner ring efficiency decreases and, the outer rings efficiency increases with increasing neutron energy. This feature is used through the ring ratio technique to determine the average energy of the neutron emission spectrum~\cite{gheorghe_2021}. Figure~\ref{fig_fed_eff_new}(b) shows the ratio of the simulated detection efficiencies of the outer and inner rings for e\-vap\-o\-ra\-tion and monochromatic spectra versus the average neutron energy. We notice that, as the $f_{RR-evap}$ e\-vap\-o\-ra\-tion and $f_{RR-mono}$ monochromatic ring ratio curves diverge starting with average neutron energies of $\sim$1~MeV, the ring ratio method is in fact sensitive to the neutron emission spectrum, which has to be properly described in order to extract accurate average neutron energy values.  

\subsection{Neutron coincidence events}

\begin{figure}[t]
\centering
\includegraphics[width=0.98\columnwidth, angle=0]{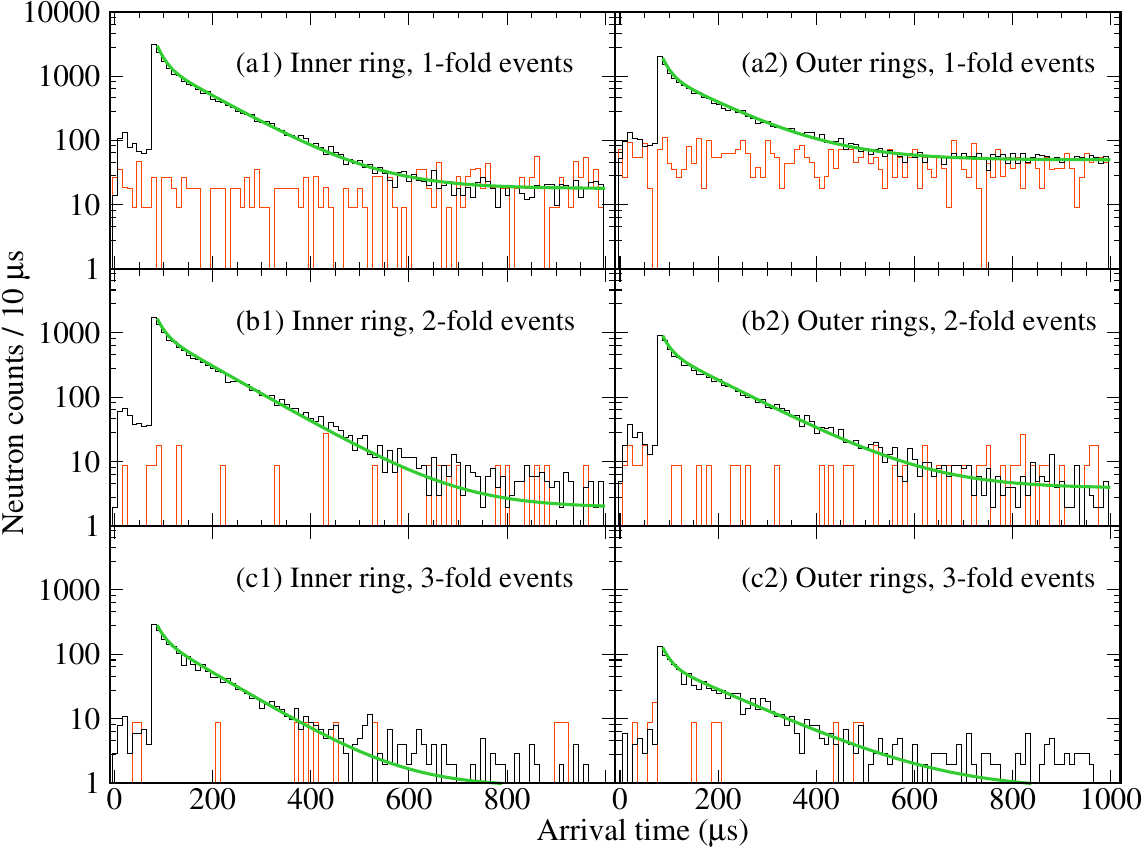} 
\caption{Arrival time distributions of neutrons emitted in photoneutron reactions on $^{208}$Pb at $E_\gamma$~=~27.72~MeV~$<$~S$_{4n}$ and recorded by the (left) inner ring and (right) summed outer rings of $^3$He counters. We plot separately the neutrons recorded in (a) 1-, (b) 2- and (c) 3-fold events. The black and red histograms correspond to beam-on and scaled beam-off data, respectively (see text). The green lines are fits to beam-on histograms.}\label{fig_ring_time}     
\end{figure}

The neutron-multiplicity sorting measurements above the two neutron separation energy $S_{2n}$ involved the recording of neutron coincidence events. By $i$-fold coincidence neutron events we refer to events where $i$ neutrons were detected during the 1~ms time interval between two consecutive $\gamma$-beam pulses. Figure~\ref{fig_ring_time} shows the histogram of the arrival time for neutrons emitted in the photoneutron reactions on $^{208}$Pb at $E_\gamma$~=~27.72~MeV~$<$~S$_{4n}$ incident photon energy. The neutron counts are discriminated by the (left) inner and (right) outer firing ring and by the coincidence order: (a) single, (b) double and (c) triple fold events. Black histograms show beam-on data and red ones show beam-off data multiplied by 9, the ratio of the measurement time with beam-on and beam-off of 90~ms and 10~ms, respectively. The flat background is due to neutrons generated by the continuous bremsstrahlung emitted by relativistic electrons circulating in the NewSUBARU ring and a small cosmic component. The background subtraction procedure discussed in Refs.~\cite{utsunomiya_2017,Gheorghe2017,gheorghe_2021} relies on fitting the time distribution with a sum of exponentials plus a flat background shown by the green lines in Fig.~\ref{fig_ring_time}. 

\section{Data analysis} \label{sec_data_analysis}

In the following, we discuss the methods for neutron multiplicity sorting, for extraction of information on the photoneutrons mean energy and spectra and on the unfolding of the measured quantities by considering the spectral distribution of the incident photon beams.

\subsection{Primary experimental quantities: $i$-fold neutron cross sections and energies}

We will here discuss the numbers of $i$-fold neutron coincidence events $n_i$ and the average energies of neutrons recorded in $i$-fold coincidences $E_i$, which are the key experimental observables necessary for extracting the photoneutron cross sections $\sigma_{inX}$ and the average neutron energies $E_{inX}$ through neutron multiplicity sorting methods.

\subsubsection{i-fold neutron cross sections: $N_i$} \label{sec_Ni_def} 

\begin{figure*}[t]
\centering
\includegraphics[width=0.98\textwidth, angle=0]{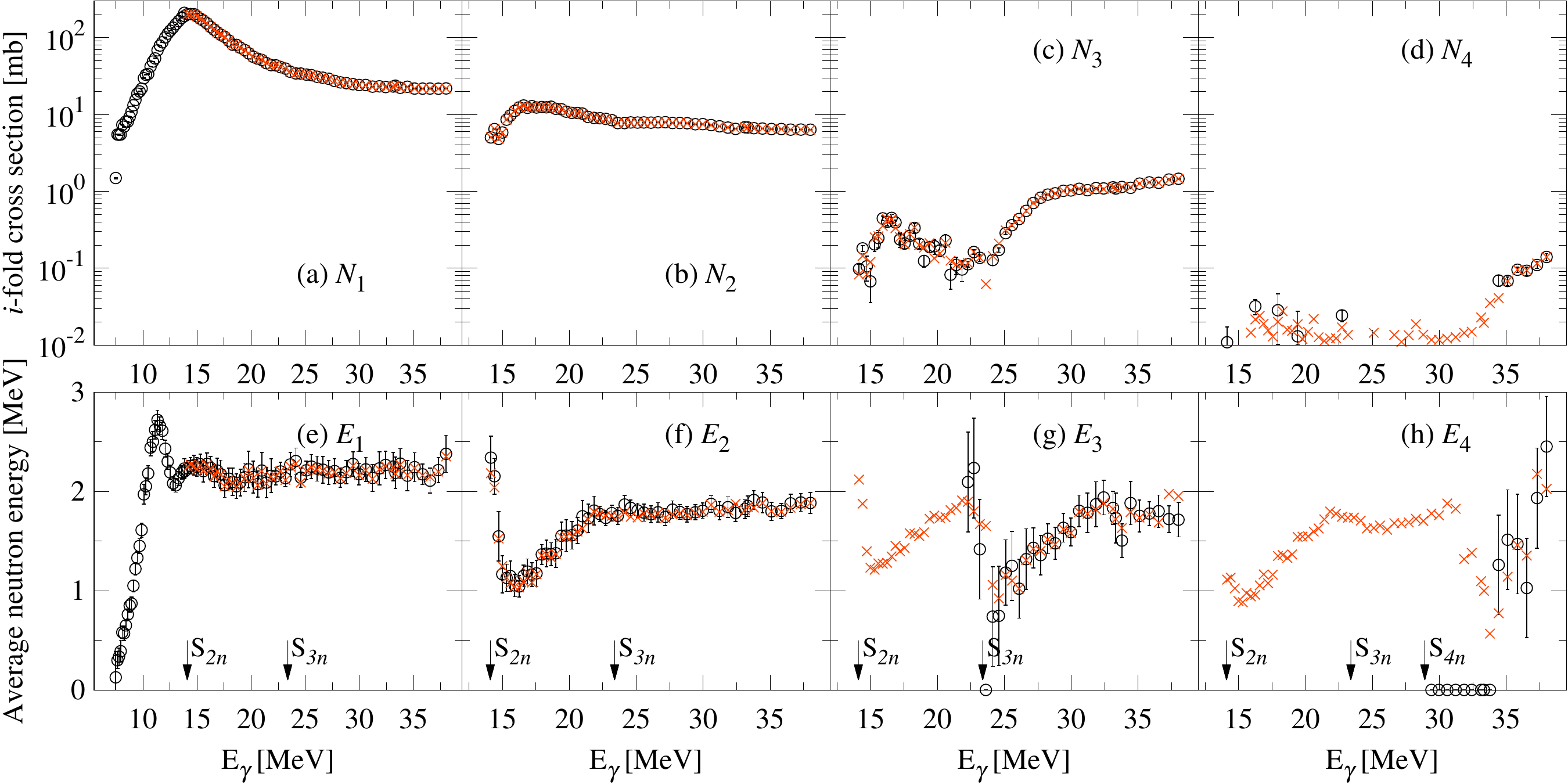} 
\caption{$^{208}$Pb $i$-fold cross sections and average neutron energies as a function of the incident photon energy. (a--d) show experimental ($N_i$, empty black dots) and best fit ($N_i^{MF}$, red crosses) $i$-fold cross sections. (e--h) show average energies of neutrons recorded in $i$-fold coincidence events (experimental - empty black dots, best fit - red crosses). The error bars for the average energies of $i$-fold coincidence events represent the statistical component and a 3$\%$ systematic component accounting for the uncertainty in the neutron detection efficiency calibration while the ones for $i$-fold cross sections $N_i$ are statistical only.}\label{fig_nev_rate}     
\end{figure*}

The number of $i$-fold neutron coincidence events $n_i$ is obtained as
\begin{equation}\label{eq_legatura_ni_rri}
n_i (E_m) = \cfrac{r_i^{out}(E_m)+r_i^{in}(E_m)}{i}, 
\end{equation} 
where $E_m$ is the maximum energy for each of the 85 incident LCS $\gamma$-ray beams used in the present experiment. $r_i^{out}$ and  $r_i^{in}$ are the numbers of neutrons recorded in $i$-fold coincidence events by the outer two rings and by the inner ring of counters, respectively. They have been obtained by integrating the corresponding background subtracted arrival time histograms.

We further introduce the $i$-fold neutron cross sections ($N_i$) for conveniently representing the $i$-fold neutron counts on the entire excitation energy range here studied, without depending on the particular target thickness and photon flux values for each experimental point.  Thus, $N_i$ is defined as: 
\begin{equation}
N_i (E_m) = \cfrac{n_i(E_m) }{N_\gamma (E_m) n_T \xi(E_m) } 
\end{equation} 
where $n_T$ is the concentration of target nuclei and $N_\gamma$ is the incident photon number. $\xi=[1-\mathrm{exp}(-\mu L)]/\mu$ is a thick target correction factor given by the target thickness $L$ and attenuation coefficient $\mu$. Figs.~\ref{fig_nev_rate}(a--d) show the experimental $i$-fold cross sections (empty black dots) for the $^{208}$Pb$(\gamma,\,inX)$ reactions with $i$ ranging from (a) 1 to (d) 4. The non-zero $N_3$ and $N_4$ values below the three and four neutron separation energies, respectively, show the presence of pile-ups, or multiple-firing neutron events which will be discussed in Sect.~\ref{sec_NMS}. 

\subsubsection{$i$-fold average neutron energies above S$_{2n}$: $E_i$} \label{sec_ei_above_s2n}

Information on the neutron spectra emitted in photoneutron reactions on $^{208}$Pb have been obtained based on the experimental $RR_i$ ring ratios for neutrons recorded in $i$-fold coincidence events, which are defined as:
\begin{equation}\label{eq_rr_exp_def}
RR_i (E_m) = \cfrac{r_i^{out}(E_m)}{r_i^{in}(E_m)}. 
\end{equation} 

For incident energies above $S_{2n}$, the average energy of neutrons recorded in $i$-fold coincidence events has been directly determined by evaluating the $f_{RR-evap}^{-1}$ neutron e\-vap\-o\-ra\-tion ring ratio function at the experimental $RR_i(E_m)$ ring ratio value (see Fig.~\ref{fig_fed_eff_new})
\begin{equation}
E_i^{exp}= f_{RR-evap}^{-1}(RR_i)\mathrm{\quad for\quad }E_m>S_{2n}.
\end{equation}
Figure~\ref{fig_nev_rate} shows the ring-ratio extracted average energies $E_i$ of neutrons recorded in $i$-fold events for i~=~1~(e) to i~=~4~(h).  

\subsubsection{$(\gamma,\,n)$ photoneutron energies below S$_{2n}$} \label{sec_e1_under_s2n}

\begin{figure}[t]
\centering
\includegraphics[width=0.98\columnwidth, angle=0]{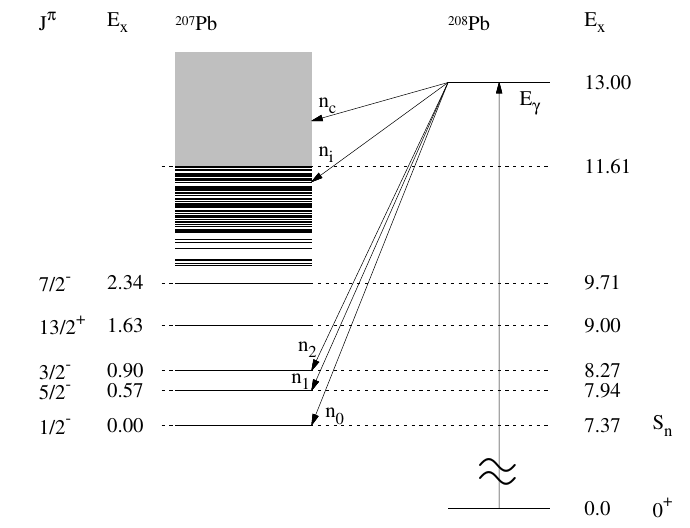} 
\caption{Partial decay scheme of $^{208}$Pb. All energy values are given in MeV.}\label{fig_schema_nivele_crop}     
\end{figure}

Figure~\ref{fig_schema_nivele_crop} shows a partial decay scheme of $^{208}$Pb illustrating the photoexcitation and decay of $^{208}$Pb to states in $^{207}$Pb by 1n-emission. E1 excitations populate $1^-$ states in $^{208}$Pb with probabilities given by the photoabsorption cross section. The $^{207}$Pb levels are well separated, with the level spacing much larger than the level width, up to excitation energies higher than 4~MeV. Based on the analysis of the cumulative number of levels, the RIPL3 recommendation \cite{capote2009_ripl} is to consider a complete level scheme of 66 discrete excited levels and start the continuum at 4.25~MeV excitation energy. 

\begin{figure}[t]
\centering
\includegraphics[width=0.98\columnwidth, angle=0]{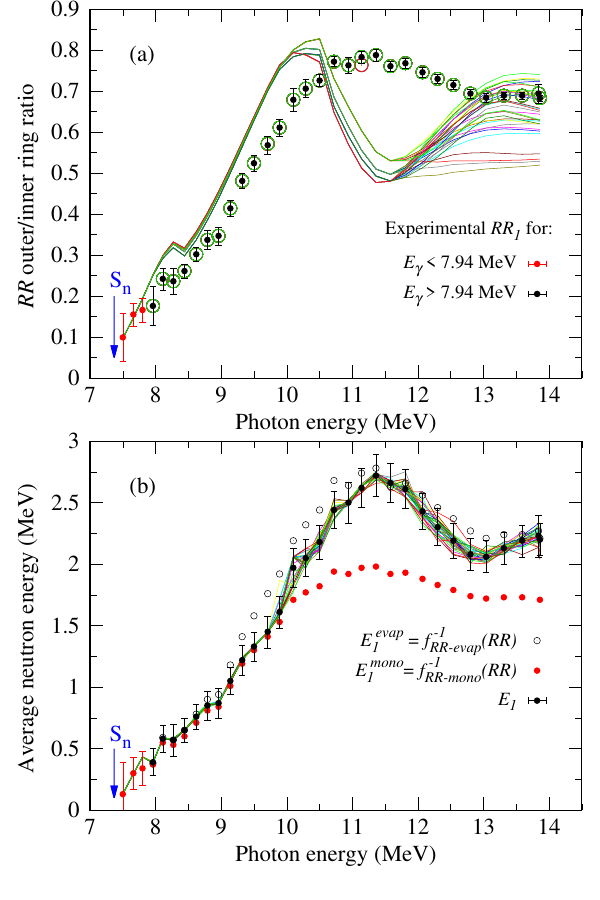} 
\caption{ $^{208}$Pb$(\gamma,\,n)$ ring ratios ($RR$) and average neutron energies ($E_1$) versus incident photon energies. (a) Experimental $RR$ (full dots) and calculated ones for the (lines) starting value and (empty dots) best-fit photoneutron spectra used in~/~resulted from the minimization procedure described in Sect.~\ref{paragraph_neutron_emission}. (b) Experimental average neutron energies obtained by reproducing the measured ring ratios (full black dots) and extracted directly through the $RR$ method assuming e\-vap\-o\-ra\-tion and, respectively, monochromatic neutron spectra. The lines show the minimization results for each set of starting values.}\label{fig_ring_ratio}     
\end{figure}

\paragraph{} For incident photon energies below $S_n$~+~0.57~MeV~=~7.94~MeV, only monochromatic neutrons can be emitted, populating the ground state of $^{207}$Pb. Such is the case for the three lowest incident energy settings at 7.50, 7.66 and 7.80~MeV, for which the experimental neutron energies have been directly determined by evaluating the $f_{RR-mono}^{-1}$ function at the measured ring ratio values $RR_1$:
\begin{equation}
E_n^{exp}= f_{RR-mono}^{-1}(RR_1)\mathrm{\quad for\quad }E_m<\mathrm{7.94~MeV}.
\end{equation}
Figures~\ref{fig_ring_ratio}(a) and (b) compare the experimental ring ratios and neutron energies (red points), respectively, with the values given by the kinematics of the two-body breakup $^{208}$Pb~$\rightarrow$~n~+~$^{207}$Pb,~$E_n=(E_\gamma-S_n)\cdot207/208$. We notice that the experimental results are consistent, within error bars, with the two-body breakup kinematics.  

\paragraph{}\label{paragraph_neutron_emission} Starting at photon beam energies larger than 7.94~MeV, neutrons can be emitted also to excited states in $^{207}$Pb. With the incident photon energy increase, the photoneutron decay transitions from emission of discrete energy neutrons to statistical emission of neutrons with continuous spectrum. 
In order to (i) correct the ring ratio extracted average neutron energies by accounting for the differences between the actual neutron emission spectra and ideal monochromatic or e\-vap\-o\-ra\-tion ones and (ii) obtain indirect information on the neutron spectra, we performed an iterative search for $^{208}$Pb$(\gamma,\,n)$ photoneutron spectra for which the $RR_1^{calc}$ calculated ring ratio values reproduce the experimental ones.

Assuming the $Y(E,E_m)$ spectrum of photoneutrons emitted at an incident energy $E_m$, the calculated ring ratio is obtained as:
\begin{equation}\label{eq_RR1calc_def}
RR_1^{calc} (E_m) = \cfrac {\sum_E Y(E,E_m) \cdot \varepsilon^{out}_{mono}(E)} {\sum_E Y(E,E_m) \cdot \varepsilon^{in}_{mono}(E)},
\end{equation}
where $\varepsilon^{in}_{mono}(E)$ and $\varepsilon^{out}_{mono}(E)$ are the inner ring and respectively, the outer rings detection efficiencies for monochromatic emission of $E$ energy neutrons. We used the $\texttt{CERN}$ $\texttt{minuit}$ package to perform a $\chi^2$ minimization procedure and determine the $Y(E,E_m)$ neutron spectrum for which the calculated ring ratio reproduces the experimental one. Each bin content was considered as a free parameter, where the spectra binning varied between 33 keV and 100 keV, depending on the incident photon energy. The minimized quantity $\chi_{E_m}^2$ was defined as:
\begin{equation}\label{eq_RR1_X2_def}
\chi_{E_m}^2 = \cfrac {(RR_1^{calc}(E_m)-RR_1(E_m))^2} {\sigma_{RR_1}^2(E_m)}.
\end{equation}
and has been calculated for each incident photon energy $E_m$ between 7.96 and 13.86~MeV. Here $\sigma_{RR_1}(E_m)$ is the statistical uncertainty for the experimental ring ratio.

The minimization procedure has been applied considering 32 sets of starting values for the neutron spectra calculated using the EMPIRE reaction code \cite{herman_2007_empire} with combinations of different nuclear inputs: 
\begin{itemize}
\item $\gamma$-ray strength function: Standard Lorentzian, Modified Lorentzian (1, 2, and 3), Enhanced Generalized Lorentzian and Simplified Modified Lorentzian models with RIPL-3 parameters \cite{capote2009_ripl}; 
\item spherical optical model potentials (OMP) for neutrons of Koning-Delaroche (RIPL IDs 2405 and 1467) and of Weisel (RIPL ID 121) and the coupled channels one of Vonach (RIPL ID 2)~\cite{capote2009_ripl}; 
\item level densities: enhanced generalized superfluid model~\cite{herman_2007_empire}, Gilbert-Cameron and microscopic HFB level densities~\cite{capote2009_ripl} with the $\tilde{a}$ parameter for the $^{207}$Pb level density varied in an interval of 80$\%$ to 100$\%$ of the RIPL-3 recommended value,
\end{itemize}
and discrete levels retrieved from the RIPL-3 database. Sect.~\ref{sec_STAT_calc} gives more details on the EMPIRE calculations. 

\begin{figure}[t]
\centering
\includegraphics[width=0.98\columnwidth, angle=0]{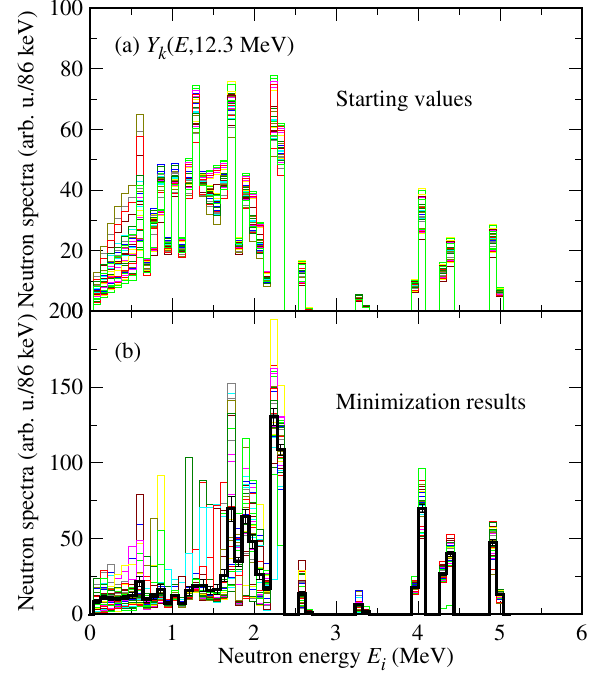} 
\caption{$^{208}$Pb$(\gamma,\,n)$ neutron spectra for 12.30~MeV incident photon energy. (a) The 32 spectra calculated with the EMPIRE statistical model code with different sets of input parameters (see text). (b) The best fit spectra obtained through $\chi^2$ minimization by reproducing the experimental ring ratio. Thin lines represent input/best fit spectra for a given set of starting values. The thick black line is the final spectrum estimation obtained as the average of the 32 best fit results.}\label{fig_sp_in_out}     
\end{figure}

Figure~\ref{fig_sp_in_out}(a) shows the EMPIRE calculations for $^{208}$Pb$(\gamma,\,n)$ neutron spectra at 12.30~MeV incident photon energy. For all sets of input parameters in the statistical model calculations, we notice the typical neutron spectrum, with a continuous component due to statistical evaporation of low energy neutrons, and discrete high energy neutron transitions due to population of low lying discrete states in the residual. The best fit neutron spectra shown in Fig.~\ref{fig_sp_in_out}(b) show that the minimization procedure systematically decreased the continuous component of low energy neutrons, thus increasing the contributions of intermediate energy neutrons of $\sim$2~MeV and of the high energy neutrons populating the low discrete states in $^{207}$Pb.

Figure~\ref{fig_ring_ratio}(a) shows the experimental ring ratios (black dots) and the ones calculated for the EMPIRE neutron spectra used as starting values in the minimization (solid lines), where each line corresponds to the one of the 32 sets of nuclear inputs. We notice significant differences between the experimental ratios and the ones computed for the starting values spectra. However, the minimization procedure identified $Y_k(E,E_m)$ best-fit photoneutron spectra for which the calculated $RR_{1,k}^{calc}$ ring ratios (empty dots) reproduce well the experimental values, where $k$ indicates the nuclear input set and cycles up to 32.  

The $E_{1,k}(E_m)$ average neutron energies were obtained as the first moments of the $Y_k(E,E_m)$ best-fit photoneutron spectra:
\begin{equation}
E_{1,k}(E_m) = \cfrac {\sum_E E \cdot Y_k(E,E_m) } {\sum_E Y_k(E,E_m)}.
\end{equation}
For each $k$ set of starting values, Fig.~\ref{fig_ring_ratio}(b) shows the $E_{1,k}(E_m)$ average neutron energies (solid lines) as function of the incident photon energy $E_m$, with each line corresponding to one of the 32 sets of minimization results. Good agreement was obtained between the average neutron energy results for each incident photon energy value. 

The final $E_1$ average energies of $^{208}$Pb$(\gamma,\,n)$ photoneutrons for excitation energies below $S_{2n}$ are shown by the full black dots in Fig.~\ref{fig_ring_ratio}(b) and have been obtained as the average of the $E_{1,k}(E_m)$ results for each incident photon energy. The error bars account for the mean standard deviation of the set of minimization results for each incident energy, the statistical uncertainty in the neutron counts for each ring of counter and a 5$\%$ uncertainty in the ring ratio average neutron energy determination.

Figure~\ref{fig_ring_ratio}(b) also compares the present $E_{1}$ average neutron energies to the ones directly extracted through the ring ratio method assuming either e\-vap\-o\-ra\-tion (empty black dots) or monochromatic (full red dots) neutron emission spectra. For incident photon energies below 8.5~MeV there is good agreement between all three curves, which follows from the overlap between the $f_{RR-evap}$ and $f_{RR-mono}$ ring ratio curves for low neutron energies below $\sim$1~MeV. At incident photon energies between 8.5 and 10~MeV, the best-fit $E_1$ values reproduce well the monochromatic ones and are well below the e\-vap\-o\-ra\-tion assumptions. For photon energies above 10~MeV, the best-fit $E_1$ results follow closely the e\-vap\-o\-ra\-tion curve, suggesting a transition to a statistical neutron emission. We notice that the ring ratio minimization procedure applied only small corrections of up to $\sim$10$\%$ to the $^{208}$Pb$(\gamma,\,n)$ average neutron energies in the transition excitation energy range between 7.94~MeV and $S_{2n}$. 

The $Y(E,E_m)$ final estimations for the $^{208}$Pb$(\gamma,\,n)$ photoneutron spectra at $E_m$ incident photon energy have been obtained by averaging each neutron energy bin content of the set of 32 best-fit $Y_k(E,E_m)$ results. This can be observed in Fig.~\ref{fig_sp_in_out}(b) for 12.30~MeV incident photon energy. The resulting $Y(E,E_m)$ neutron spectra are shown and discussed in Sect.~\ref{sec_results} along with the partial cross sections for the $^{208}$Pb$(\gamma,\,n)$ reactions that populate the residual nucleus $^{207}$Pb in its ground state and in the first and second excited states. 

\subsection{Neutron multiplicity sorting} \label{sec_NMS}

Let us consider a photon beam of energy $E_\gamma$ incident on the $^{208}$Pb target, where $S_{\texttt{N}n}<E_\gamma < S_{(\texttt{N}+1)n}$ and $S_{\texttt{N}n}$ is the separation energy of $\texttt{N}$ neutrons, which are given in Table~\ref{table_Sxn} for neutron multiplicities up to 5. Photoneutron $(\gamma,\,inX)$ reactions with $i$ cycling up to $\texttt{N}$ will be induced in the target, each characterized by the $\sigma_{inX}$ cross section and by the $E_{inX}$ average energy of the photoneutron spectrum. For (i) $single$-firing conditions in which no more than one reaction is induced in the target by the same photon pulse, and (ii) constant neutron detection efficiency, the Direct Neutron Multiplicity (DNM) sorting method relates the $\sigma_{inX}$ cross sections to the raw $N_i$ neutron coincidence experimental observables as~\cite{utsunomiya_2017}:
\begin{equation} \label{EQ_DNM}
N_i = \sum_{x=i}^{\texttt{N}} \sigma_{xnX} \cdot {}_{x}C_{i} \varepsilon^i (1-\varepsilon)^{x-i},
\end{equation}
and the $E_{inX}$ photoneutron energies to the $E_i$ average energies of neutrons detected in $i$-fold coincidences as~\cite{gheorghe_2021}:
\begin{equation} \label{EQ_E_k}
E_i = \sum_{x=i}^\texttt{N} E_{xnX} \sigma_{xnX} \cdot {}_{x}C_{i} \varepsilon^i (1-\varepsilon)^{x-i} / N_i, 
\end{equation}
where $\varepsilon$ is the constant neutron detection efficiency and  ${}_{x}C_{i}$ is the binomial coefficient. 

\begin{table}[b]
\caption{\label{table_Sxn} $^{208}$Pb separation energies for $i$ neutrons $S_{in}$ in MeV.}
\begin{ruledtabular}
\begin{tabular}{lccccc}
   \textrm{S$_n$} & \textrm{S$_{2n}$}& \textrm{S$_{3n}$} & \textrm{S$_{4n}$} & \textrm{S$_{5n}$} \\ \colrule
           7.368  &  14.107          & 22.194            & 28.927            & 37.323 \\
\end{tabular}
\end{ruledtabular}
\end{table}

For $single$-firing conditions to be met, low reaction rates were maintained by conveniently limiting the target thickness and incident photon flux. Still, as the number of photons in the LCS $\gamma$-ray pulses followed Poisson distributions with mean values between 6 and 20 photons per pulse, the incidence of $multiple$-firings has only been minimized, but not completely eliminated. Multiple firings are visible in Figs.~\ref{fig_nev_rate}(c) and (d) from the non-zero $N_3$ and $N_4$ cross sections, respectively, below the three neutrons separation energy $S_{3n}$. Explicitly, the 3-neutron coincidence events below $S_{3n}$ originate from $(\gamma,\,n)$ and $(\gamma,\,2n)$ reactions induced by the same $\gamma$-ray pulse. 

Thus, we applied the statistical treatment of neutron coincidence events described in Ref.~\cite{gheorghe_2021}, which models the $multiple$-firing of all available combinations of photoneutron $(\gamma,\,inX)$ reactions with corresponding contributions given by the reaction cross sections, areal density of the irradiated samples and the multiplicity of incident photons per pulse. Using the $\texttt{CERN}$ $\texttt{minuit}$ package, a $\chi^2$ minimization procedure has been performed to determine the $multiple$-firing corrected $^{208}$Pb cross sections and average photoneutron energies from the measured  $i$-fold cross sections $N_i$ and average energies $E_i$. The $^{208}$Pb$(\gamma,\,inX)$ cross sections and average neutron energies are the free parameters in the minimization procedure, while the target ($n_T$, $\xi$) and LCS $\gamma$-ray beam ($N_\gamma$, mean number of photons per pulse) characteristics have been fixed to the experimentally determined values. The best-fit $i$-fold cross sections $N_i^{M\!F}$ and average neutron energies $E_i^{M\!F}$ are shown in Fig.~\ref{fig_nev_rate} in comparison with the raw measured values. We notice that the raw measured data are well reproduced by the $multiple$-firing results on the entire excitation energy region and for all observed neutron multiplicities. 

\begin{figure*}[t]
\centering
\includegraphics[width=0.98\textwidth, angle=0]{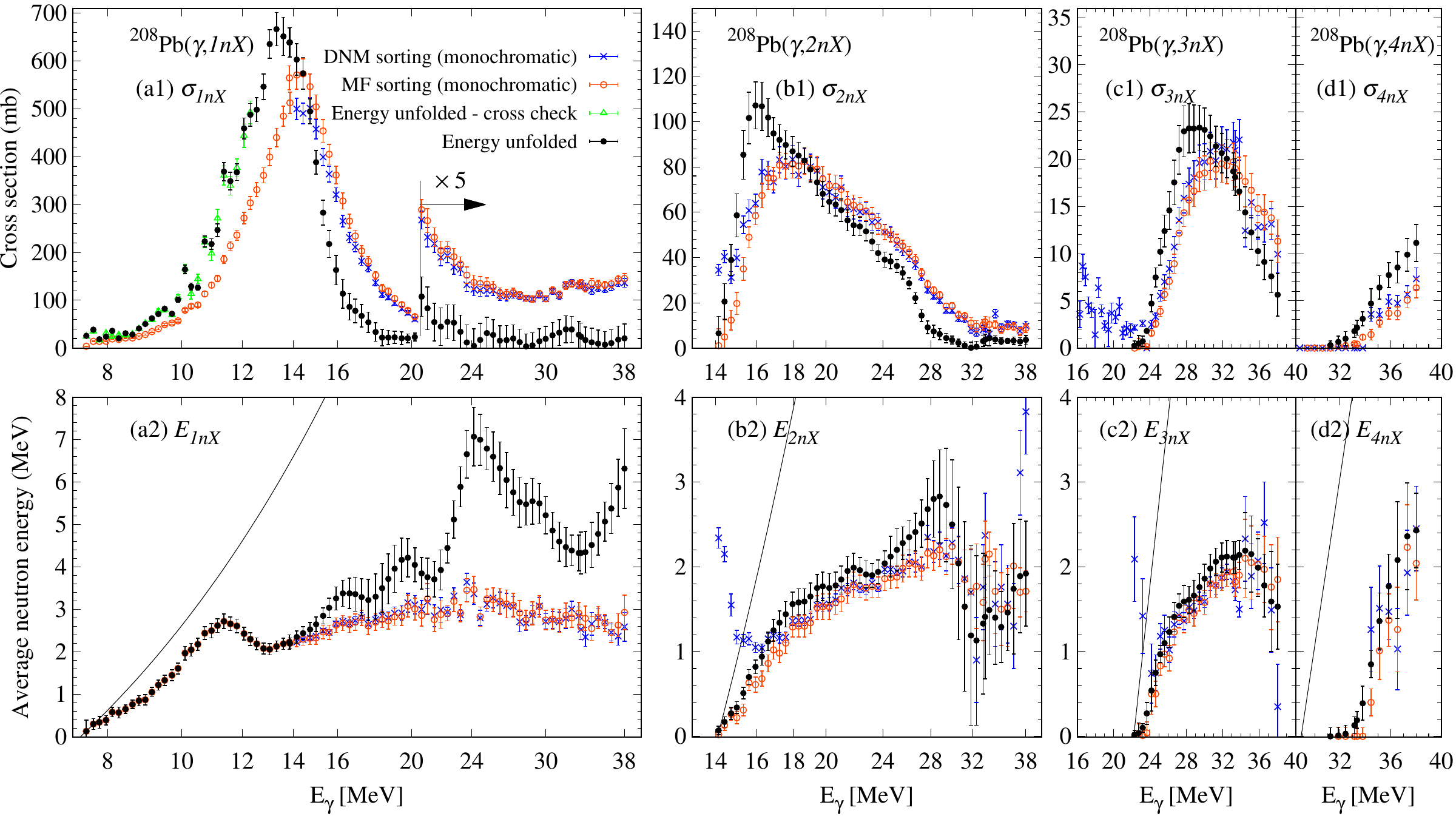} 
\caption{Present $^{208}$Pb results in monochromatic approximation as obtained with the DNM (blue crosses) and with the $multiple$-firing (MF) (empty red dots) sorting methods compared with the energy unfolded results (full black dots): (upper figures (a--d)1) cross sections of and (lower figures (a--d)2) average energies of neutrons emitted in the (a)~$(\gamma,\,1nX)$, (b)~$(\gamma,\,2nX)$, (c)~$(\gamma,\,3nX)$ and (d)~$(\gamma,\,4nX)$ reactions. The low energy $\sigma_{\gamma,\,n}$ unfolding has been cross checked by two independent methods (see text). The error bars account for the total uncertainty. The solid lines correspond to the maximum neutron energies given by kinematics $208/207\cdot(E_\gamma-S_{in})$. The (a) and (b) horizontal axes are in log scale.}\label{fig_mono_vs_unfolded}  
\end{figure*}

Figure~\ref{fig_mono_vs_unfolded} shows the present neutron multiplicity sorting results for the $^{208}$Pb$(\gamma,\,inX)$ reactions with (a)~$i$~=~1, (b)~$i$~=~2, (c)~$i$~=~3 and (d)~$i$~=~4. The DNM cross sections $\sigma_{inX}^{D\!N\!M}$ and average photoneutron energies $E_{inX}^{D\!N\!M}$ obtained by solving the systems of equations~\eqref{EQ_DNM}, and \eqref{EQ_E_k}, respectively, are shown by the blue crosses. The $multiple$-firing (MF) corrected ones $\sigma_{inX}^{M\!F}$ and $E_{inX}^{M\!F}$ are shown by the empty red dots and correspond to the best-fit $N_i^{M\!F}$ and $E_i^{M\!F}$ shown in Fig.~\ref{fig_nev_rate}. 

\paragraph{$\sigma_{inX}$ results.} The reasonable fulfillment of the $single$ firing conditions is indicated by the increasing behavior of the $\sigma^{D\!N\!M}_{inX}$ with i~=~2,~3,~4 starting from the reactions threshold and their drop with the opening of the channel with the immediately higher neutron emission multiplicity. Still, $multiple$-firing corrections have produced visible changes to the cross sections, especially in the excitation energy regions just above $S_{in}$ reaction thresholds, bringing a general improvement in the scatter of the cross sections for all channels.  The $(\gamma,\,2n)$ cross sections in the vicinity of $S_{2n}$ have been decreased by taking into account the multiple firings of the strong $(\gamma,\,1nX)$ channel. The non-zero $\sigma^{D\!N\!M}_{3nX}$ and $\sigma^{D\!N\!M}_{4nX}$ values below $S_{3n}$ were reallocated to the $(\gamma,\,1nX)$ and $(\gamma,\,2nX)$ channels. 

\paragraph{$E_{inX}$ results.} The $multiple$-firing corrections applied to the $E_{inX}^{D\!N\!M}$ energies are generally small and within error bars. Exceptions are seen in the incident energy regions just above the $S_{2n}$ and $S_{3n}$ thresholds, where the weak newly opened $(\gamma,\,2n)$ and $(\gamma,\,3n)$ channels have small photoneutron energies compared with the stronger channels of lower neutron emission multiplicities. Explicitly, the $multiple$ firings of $(\gamma,\,1nX)$ reactions with high $E_{1nX}$ values artificially increase the average energy of double neutron events to values higher than the maximum energies given by kinematics $207/208\cdot(E_\gamma-S_{2n})$. We notice that the average neutron energies show strong statistical fluctuations at high excitation energies, which confirms the need of a flat efficiency neutron detector for precise reaction cross section measurements. 

\subsection{Energy unfolding}

The $(\gamma,\,inX)$ cross sections and average photoneutron energies obtained through the neutron multiplicity sorting method discussed above are referred to as monochromatic approximations representing in fact the folding of the true, energy-dependent quantities with the spectral distribution of the incident photon beams. The measured cross sections are given by the folding of the excitation function $\sigma_{inX}(E_\gamma)$ with the beam spectral distribution $L(E_\gamma,E_m)$:
\begin{equation} \label{eq_folded_cs}
\sigma_{inX}^{M\!F}(E_m)= \cfrac{1}{\xi} \int_0^{E_m} L(E_\gamma,E_m) \sigma_{inX}(E_\gamma)\,dE_\gamma,
\end{equation}
where $\xi$ is the thick target correction factor introduced in Sec.~\ref{sec_Ni_def}. The measured average energies are obtained by folding the incident energy-dependent function $E_{inX}(E_\gamma)$ with the beam spectral distribution and the excitation function:
\begin{equation} 
E_{inX}^{M\!F}(E_m)= \cfrac{\int_0^{E_m} \!\! E_{inX}(E_\gamma) L(E_\gamma,\!E_m)\sigma_{inX}(E_\gamma)dE_\gamma}{\sigma_{inX}^{M\!F}(E_m)\xi}. \label{eq_folded_en_gxn}
\end{equation}
The $\sigma_{inX}^{M\!F}(E_m)$ and $E_{inX}^{M\!F}(E_m)$ monochromatic approximations are not connected to a specific incident photon energy $E_\gamma$, but are functions of the LCS $\gamma$-beams maximum energy $E_m$. 

An iterative unfolding procedure described in Refs.~\cite{Renstrom18,LarsenTveten23} has been applied to the quantities defined in Eqs.~\eqref{eq_folded_cs}~and~\eqref{eq_folded_en_gxn} in order to extract the energy-dependent photoneutron cross sections and average energies. The method was applied independently for each quantity $\sigma_{inX}(E_\gamma)$ and $E_{inX}(E_\gamma)$ with $i$~=~1~to~4, first for the cross sections and then for the average neutron energies.

For each unfolding process, the method starts with a constant trial function of 10 keV binning spanning from the corresponding $S_{in}$ threshold up to the maximum investigated energy of 38.02~MeV. The trial function is iteratively adjusted based on the difference between its folding with the beam spectral distribution and the measured quantity. The iterations stop when the folded trial function reproduces the measured cross sections (average neutron energies). The adjustment of the trial function has been performed independently for each energy bin using linear interpolations between the measured/folded values as functions of $E_m$. To prevent spurious fluctuations being introduced by the unfolding method, after each iteration we applied an energy-dependent (0.1~--~1.4)~MeV smoothing factor related to the FWHM energy distribution of the LCS $\gamma$-ray beams listed in Table~\ref{table_beams}. 

Particular attention has been paid to the unfolding of the resonant $\sigma_{\gamma,\,n}$ cross sections in the energy excitation region below $S_{2n}$. Here, we found that smoothing factors smaller than the experimental FHWM spectral distributions were needed in order to reproduce the rapid changes in the monochromatic approximation of the $(\gamma,\,n)$ cross section. To validate the results, we performed a separate unfolding process in which we described the photoabsorption cross section as a sum of analytical functions: two Lorentz functions for the entire GDR region and eight Gauss functions for describing the resonant structure below $E_\gamma$~=~13~MeV. Using the $\texttt{CERN}$ $\texttt{minuit}$ package, we searched for an excitation function that, folded with the $\gamma$ beam spectral distributions, optimally reproduced the monochromatic $(\gamma,\,n)$ cross sections below $S_{2n}$. The strength, centroid, and width of the Lorentz and Gauss functions were considered as free parameters in the minimization. The resulting $\sigma_{\gamma,\,n}$, shown by the empty green triangles in Fig.~\ref{fig_mono_vs_unfolded}(a1), reproduced within error bars the cross sections obtained through the iterative method of Ref.~\cite{Renstrom18}. Finally, we adopted the results obtained by the iterative method of Ref.~\cite{Renstrom18}, which makes no assumptions on the shape of the unfolded cross sections. 

Figure~\ref{fig_mono_vs_unfolded} shows the energy unfolded cross sections and average neutron energies (full black dots) along with the monochromatic approximation values. The spectral distribution of the incident $\gamma$-ray beams has already been considered in the procedure applied for extracting the $E_1$ average energy of $(\gamma,\,n)$ neutrons, and thus the unfolded values (black dots) coincidence with the ones obtained in Sect.~\ref{sec_e1_under_s2n} (red empty dots). The error bars for the energy unfolded results account for the statistical uncertainties in the neutron detection and for uncertainties of 3$\%$ for the neutron detection efficiency~\cite{utsunomiya_2017,Gheorghe2017}, 3$\%$ for the photon flux determination, 1$\%$ for the target thickness and the incident photon spectra uncertainty. The uncertainty in the incident photon spectra, shown by the red and green bands in Fig.~\ref{fig_inc_spectra_lin2}, has been propagated by applying the unfolding procedure separately for the upper and lower limit of the incident spectra.

\section{Experimental results}\label{sec_results}

\subsection{Photoabsorption cross sections}

\begin{figure}[t]
\centering
\includegraphics[width=0.98\columnwidth, angle=0]{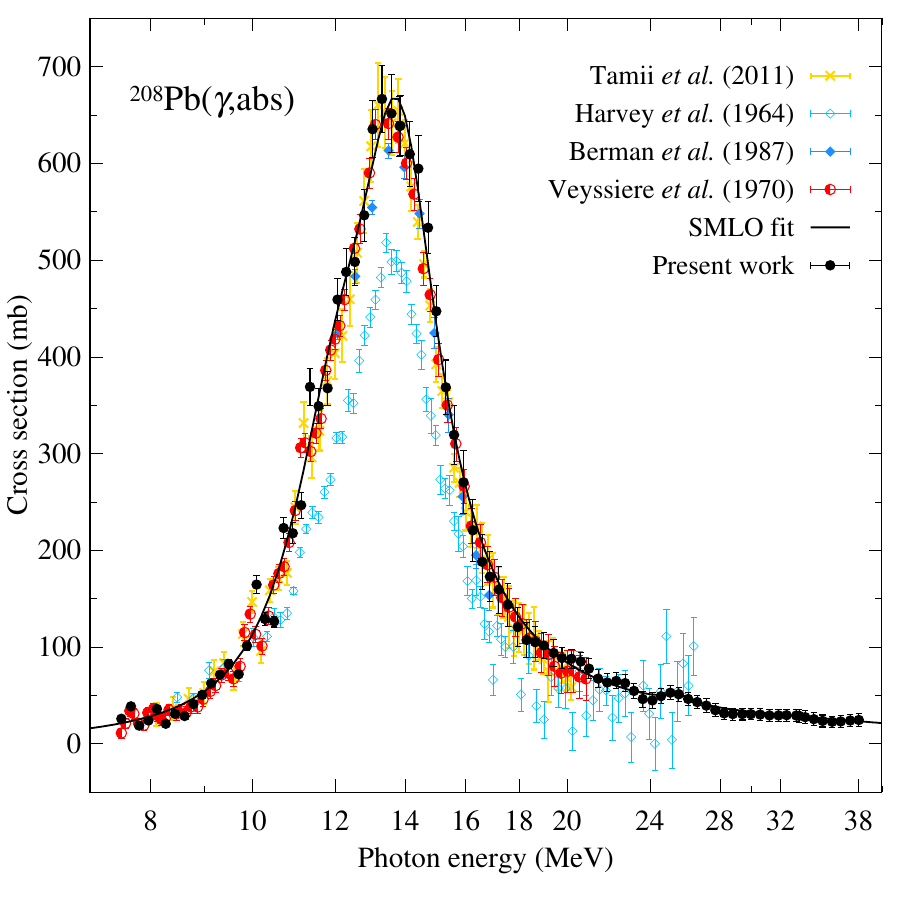} 
\caption{Present photoabsorption cross sections for $^{208}$Pb (full dots) compared with existing positron in flight annihilation data of Saclay~\cite{veyssiere_1970} (half full dots) and Livermore~\cite{harvey_1964,berman_1987} (diamonds) and more recent, indirect data obtained in hadronic experiments at RCNP Grand~Raiden~\cite{tamii_2011} (crosses). The SMLO curve corresponds to a three-Lorentzian fit to the present data using the Simple Modified Lorentzian function described in Ref.~\cite{plujko_2018,goriely_2019}, as discussed in Sect.~\ref{sec_STAT_calc}. The present results in numerical format are available in Ref.~\cite{supplemental_material}.}\label{fig_cs_abs}     
\end{figure}

Based on the negligible contributions of $(\gamma,\,p)$, $(\gamma,\,\alpha)$, etc. charged particle only emission reactions, which are highly suppressed by the large Coulomb barrier, the total photoabsorption cross section in $^{208}$Pb is well approximated by the sum cross section for the neutron emission reactions:
\begin{align} \label{eq_abs_cs}
\sigma_{abs} & \equiv \sigma_{\gamma, \,abs} \nonumber \\
             & \approx \sigma_{1nX} + \sigma_{2nX} + \sigma_{3nX} + \sigma_{4nX}
\end{align}
The so obtained $^{208}$Pb photoabsorption cross sections are shown in Fig.~\ref{fig_cs_abs} in comparison with preceding data. The present results are in good agreement with the Saclay data of Veyssiere~$et$~$al.$~\cite{veyssiere_1970} and with the indirectly determined photoabsorption cross sections extracted from RCNP proton inelastic scattering experiments by Tamii~$et$~$al.$~\cite{tamii_2011}. The Livermore cross sections obtained by Harvey~$et$~$al.$~\cite{harvey_1964} for the entire GDR energy range strongly underestimate all four other sets of results shown in Fig.~\ref{fig_cs_abs}. The cross sections remeasured by Berman~$et$~$al.$~\cite{berman_1987} at Livermore using a natural lead sample are only slightly lower than the present results, by $\sim$5$\%$ for the GDR peak value. We note instead that the centroid and width of the giant dipole resonance is in good agreement for all experimental data sets.

\subsection{Photoneutron cross sections}

\begin{figure}[t]
\centering
\includegraphics[width=0.98\columnwidth, angle=0]{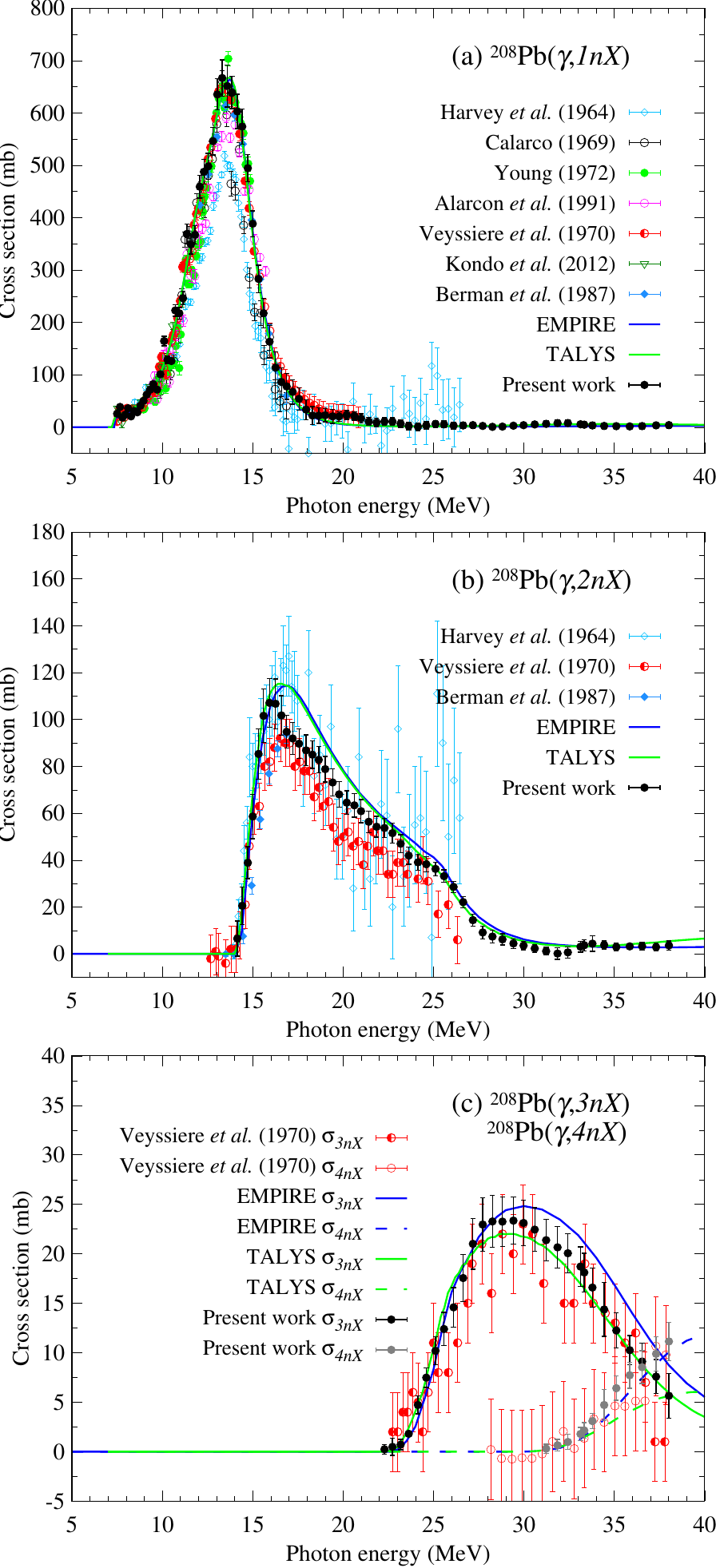} 
\caption{Present photoneutron cross sections for the (a)~$(\gamma,\,1nX)$, (b)~$(\gamma,\,2nX)$, (c)~$(\gamma,\,3nX)$ and $(\gamma,\,4nX)$ reactions in $^{208}$Pb compared with existing data and EMPIRE and TALYS statistical model calculations.}\label{fig_cs_gxn}     
\end{figure}

Figure~\ref{fig_cs_gxn}(a) shows the present $\sigma_{1nX}$ photoneutron cross sections for $^{208}$Pb compared with previous measurements. As for the photoabsorption, we notice a good agreement with the Saclay data of Veyssiere~$et$~$al.$~\cite{veyssiere_1970}, a strong underestimation of the present results by the Livermore data of Harvey~$et$~$al.$~\cite{harvey_1964} and $\sim$5$\%$ higher present GDR peak cross section values than the Berman~$et$~$al.$~\cite{berman_1987} data. The present $\sigma_{1nX}$ are also in good agreement the recent measurements performed by Kondo~$et$~$al.$~\cite{kondo_2012} (green empty triangles) using LCS $\gamma$-ray beams with energies up to 13~MeV and a moderated neutron detection array of $^3$He counters. Good agreement with bremmstrahlung monochromator data is observed for the unpublished results of Young~\cite{young_1972}~(full green dots) in the GDR peak region and with those of Calarco~\cite{calarco_1969} only on the increasing slope of the $\sigma_{1nX}$. The $\sigma_{1nX}$ cross sections of Alarcon~$et$~$al.$~\cite{alarcon_1991} (magenta empty diamonds), which were obtained by integrating differential cross sections over neutron energy and angles, show GDR values lower than the present data. 

The present $^{208}$Pb$(\gamma,\,2nX)$ cross sections shown in Fig.~\ref{fig_cs_gxn}(b) are systematically higher than the Veyssiere~$et$~$al.$~\cite{veyssiere_1970} and Berman~$et$~$al.$~\cite{berman_1987} ones. A better agreement is found with the Harvey~$et$~$al.$~\cite{harvey_1964} results below the $\sim$18~MeV. Above $\sim$18~MeV, it is difficult to make a comparison because of the large statistical fluctuations of the Livermore results. 

Figure~\ref{fig_cs_gxn}(c) shows the present $^{208}$Pb$(\gamma,\,3nX)$ and $^{208}$Pb$(\gamma,\,4nX)$ cross sections, which are in good agreement with the Saclay results in the excitation energy region up to 38~MeV. 

The present photoneutron cross sections are available in numerical format in Ref.~\cite{supplemental_material}. The EMPIRE and TALYS calculations shown in Fig.~\ref{fig_cs_gxn} by the blue and green lines, respectively, are discussed in Sect.~\ref{sec_STAT_calc}.

\subsection{Resonant structures in low-energy $^{208}$Pb$(\gamma,\,n)$ cross sections}

\begin{figure}[t]
\centering
\includegraphics[width=0.98\columnwidth, angle=0]{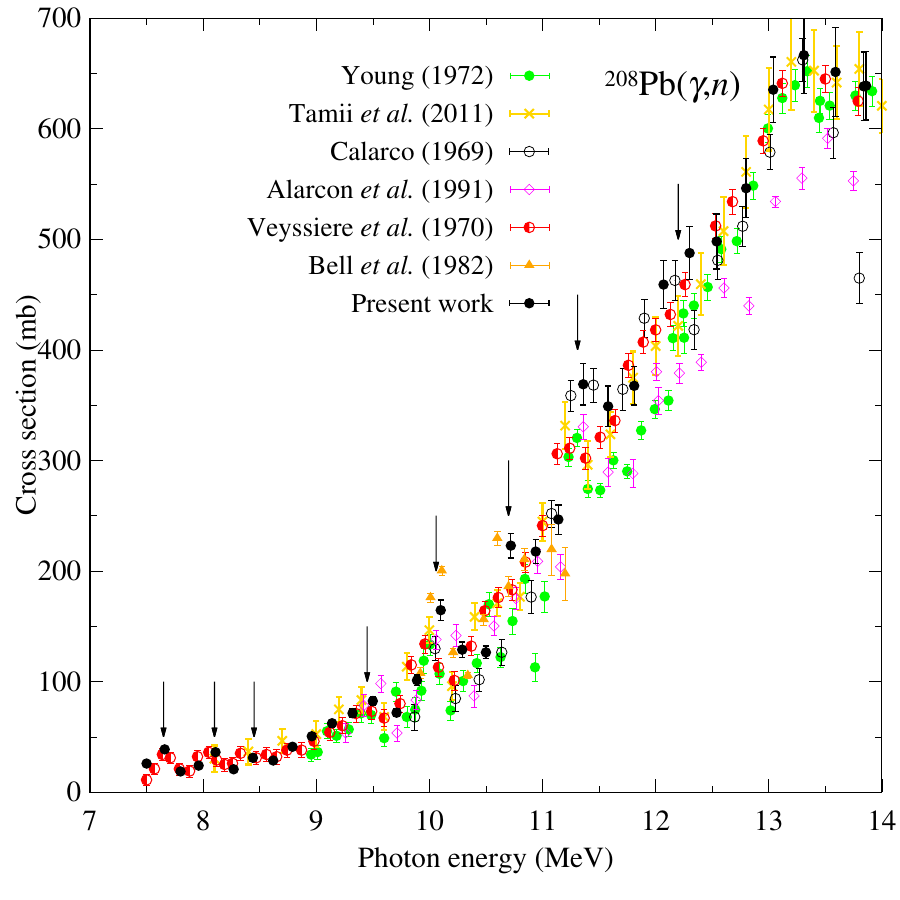} 
\caption{Present low energy $^{208}$Pb$(\gamma,\,n)$ cross sections compared with existing high energy resolution measurements. Arrows indicate resonance structures.}\label{fig_cs_g1n_res}     
\end{figure} 

Figure~\ref{fig_cs_g1n_res} shows the $^{208}$Pb$(\gamma,\,n)$ cross sections in the low excitation region below the $S_{2n}$ threshold. The present cross sections are plotted along with existing high resolution measurements in order to compare the results on the resonant structures below $E_\gamma$~=~13~MeV. The peaks at 7.65, 8.10 and 8.45~MeV reproduce well the structures observed in the Saclay experiment~\cite{veyssiere_1970}. The broad shoulder at 9.45~MeV has also been observed both in the Saclay and RCNP~\cite{tamii_2011} experiments and in the tagged bremmstrahlung measurements of Young~\cite{young_1972} and Alarcon~$et$~$al.$~\cite{alarcon_1991}. For the prominent $\sim$10~MeV peak we obtained a 10.06~MeV centroid value, which is in agreement with the tagged bremmstrahlung experiments of Bell~$et$~$al.$~\cite{bell_1982} and Alarcon, but about $\sim$150~keV higher than the peak position obtained in the Saclay experiment. The 10.70~MeV peak has been previously observed only in the Bell measurements, while the Saclay and RNCP experiments described it as a low shoulder. The present 11.31~MeV peak position confirms the tagged bremmstrahlung results of Alarcon and Calarco~\cite{calarco_1969}, but is again $\sim$100~keV higher than the Saclay and RCNP centroid values. The 12.20~MeV shoulder is also observed in all the other measurements. 

\subsection{Average photoneutron energies}

\begin{figure}[t]
\centering
\includegraphics[width=0.96\columnwidth, angle=0]{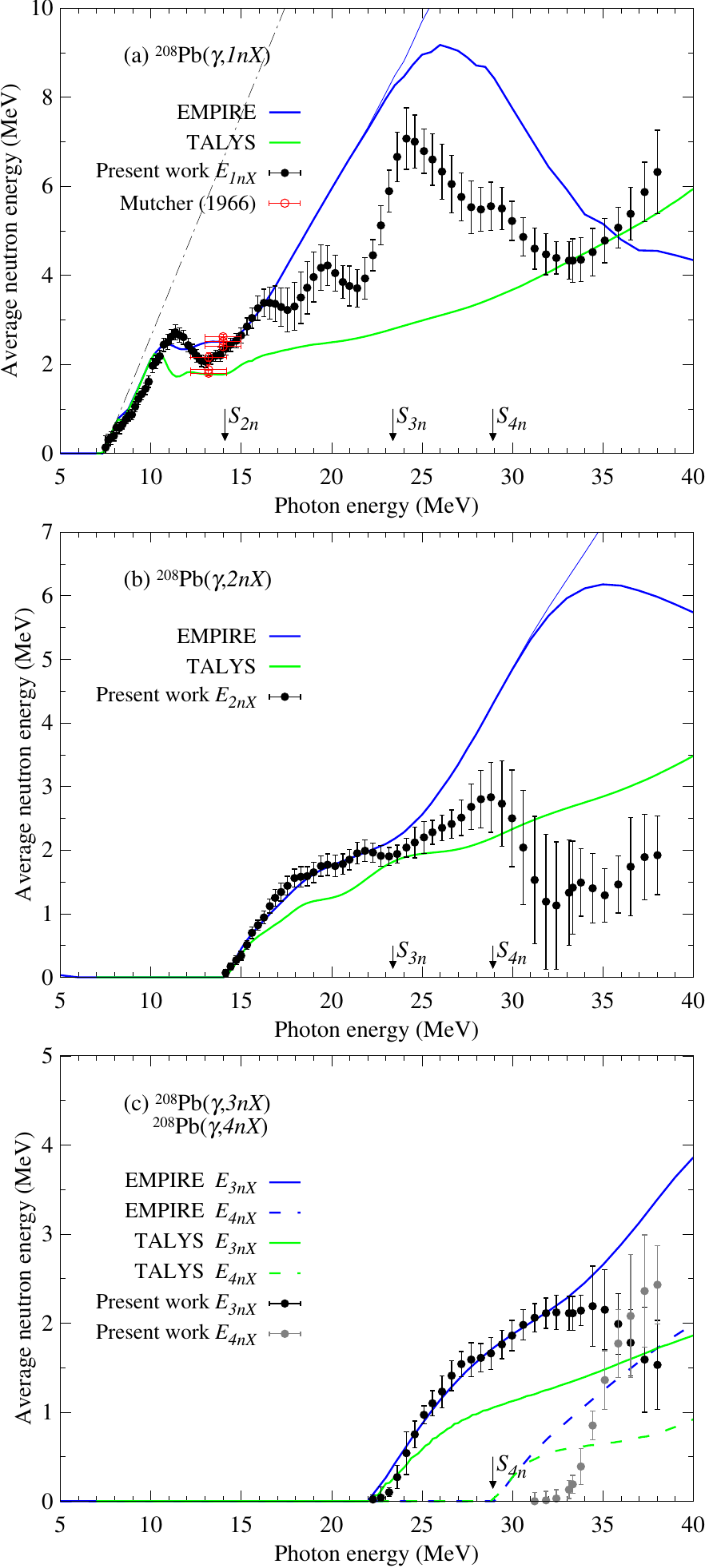} 
\caption{Present photoneutron average energies for the (a)~$(\gamma,\,1nX)$, (b)~$(\gamma,\,2nX)$, (c)~$(\gamma,\,3nX)$ and $(\gamma,\,4nX)$ reactions in $^{208}$Pb compared with EMPIRE and TALYS calculations. Thin blue lines show EMPIRE $(\gamma,\,in)$ average energies.}\label{fig_nen_gxn}     
\end{figure}

Figure~\ref{fig_nen_gxn} shows the average energies of photoneutron energies emitted in the (a)~$(\gamma,\,1nX)$, (b)~$(\gamma,\,2nX)$, (c)~$(\gamma,\,3nX)$ and $(\gamma,\,4nX)$ reactions in $^{208}$Pb. 

The energy of the $(\gamma,\,1nX)$ photoneutrons shows a steep increase starting from $S_n$ up to 11.5 MeV excitation energy followed by a fast drop until the opening of the $(\gamma,\,2n)$ channel. Above $S_{2n}$, the average neutron energy resumes its increase, which confirms the unpublished results of Mutchler~\cite{mutchler_1966} obtained by the photon difference method using bremsstrahlung photon beams of 13, 14 and 15~MeV and time of flight measurements at 3 angles relative to the photon beam. We notice that the average neutron energies steeply increase around 22~MeV excitation energy. 

\begin{figure*}[t]
\centering
\includegraphics[width=0.98\textwidth, angle=0]{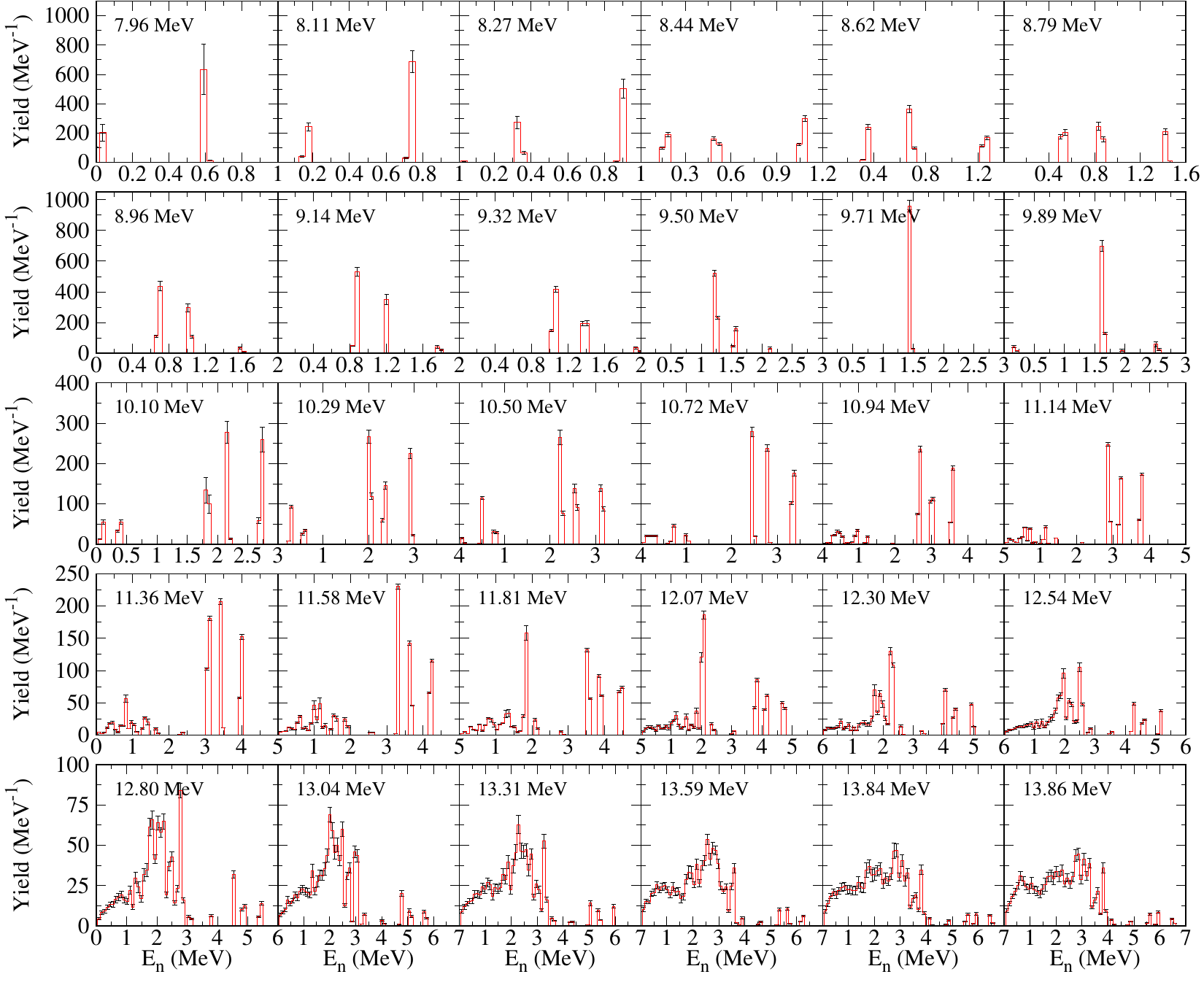} 
\caption{Present estimations for the $^{208}$Pb$(\gamma,\,n)$ photoneutron spectra normalized to an integral of 1000 counts for incident photon energies up to 13.86~MeV. }\label{fig_neutron_spectra_min_avg}     
\end{figure*} 

The $(\gamma,\,2nX)$ reaction is characterized by a long rise in the average photoneutron energy $E_{2nX}$, which increases quickly above $S_{2n}$. The growth slows down at $\sim$18~MeV excitation energy and continues until the $S_{4n}$ threshold, above which the experimental $E_{2nX}$ shows a slow decrease with strong fluctuations which follow from the statistical scatter in the unfolded average neutron energy values shown in Fig.~\ref{fig_mono_vs_unfolded}(b2). The $E_{3nX}$ average neutron energies, starting from $S_{3n}$, increase up to $\sim$5~MeV above $S_{4n}$, where they start decreasing.The $(\gamma,\,4nX)$ reaction shows a rather abrupt increase in the average neutron energy values, starting $\sim$4~MeV above the reaction threshold.
 
The present photoneutron average energies are available in numerical format in Ref.~\cite{supplemental_material}. The comparison with reaction model codes is discussed in Sect.~\ref{sec_STAT_calc}.

\subsection{Estimations for $(\gamma,\,n)$ photoneutron spectra and partial cross sections}

\subsubsection{$^{208}$Pb$(\gamma,\,n)$ photoneutron energy spectra}

Figure~\ref{fig_neutron_spectra_min_avg} shows the present estimations for the $^{208}$Pb$(\gamma,\,n)$ photoneutron spectra $Y(E,E_m)$ obtained by reproducing the experimental ring ratio values. The spectra represent branching ratios normalized so that their integral is equal to 1000. The error bar for each energy bin content is the quadratic sum of the ring ratio statistical error and of the standard deviation of the mean over $Y_k(E,E_m)$, with $k$ indicating the set of nuclear inputs used for the minimization starting values. 

The neutron emission to the ground (0~keV, 1/2$^-$), first (569.7~keV, 5/2$^-$) and second (897.7~keV, 3/2$^-$) excited states, which are recognized by the three highest energy discrete neutron emissions, show significant yields up to incident photon energies of $\sim$13~MeV. Starting with incident photon energies above 10~MeV, we notice significant population yields for the higher excited states. This is in agreement with the behavior of the average energy of the total $^{208}$Pb$(\gamma,\,n)$ photoneutron spectra, which, as shown in Fig.~\ref{fig_ring_ratio}(b) follow closely the e\-vap\-o\-ra\-tion ring ratio average energy curve for excitation energies higher than 10~MeV. We also notice that the present estimations obtained by reproducing the experimental ring ratio values confirm the transition to a statistical neutron emission towards the $(\gamma,\,2n)$ reaction threshold. 

\subsubsection{Partial photoneutron cross sections for $E_\gamma < S_{2n}$}

\begin{figure}[t]
\centering
\includegraphics[width=0.98\columnwidth, angle=0]{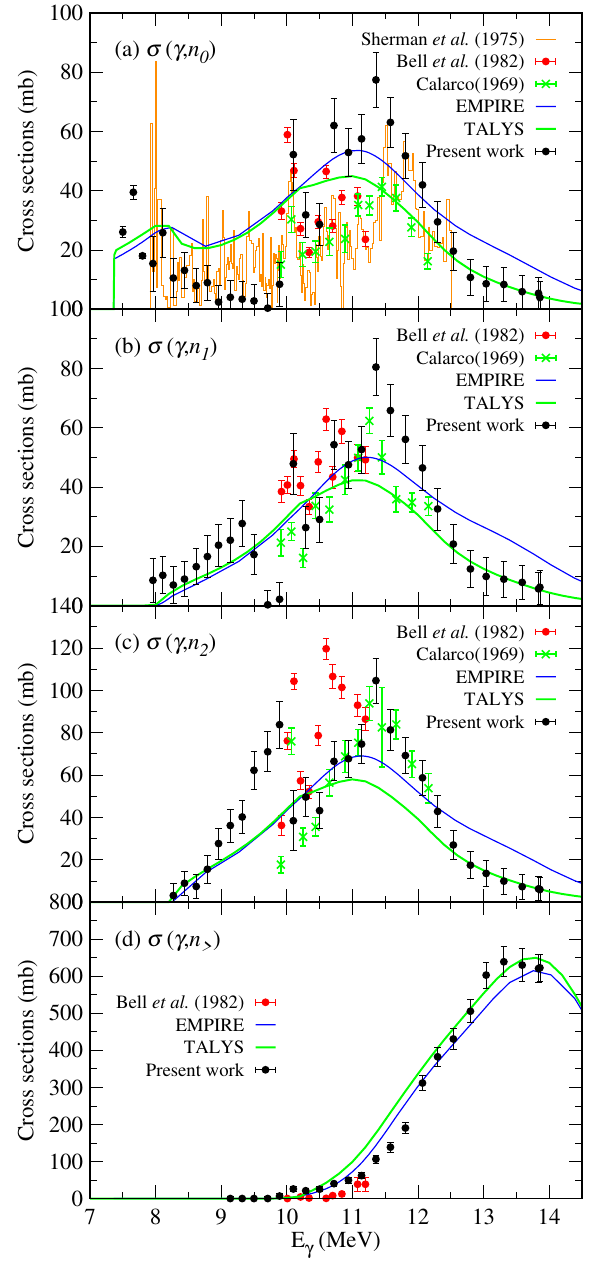} 
\caption{Partial cross sections for the $^{208}$Pb$(\gamma,\,n)$ reaction leaving $^{207}$Pb in (a) its ground state, in its (b) first and (c) second excited states and (d) in any excited state higher than the second. Present estimations are compared with existing data. Present results are given in numerical format in Ref.~\cite{supplemental_material}.}\label{fig_cs_partiale}     
\end{figure}

Figure~\ref{fig_cs_partiale} shows the partial cross sections for $^{208}$Pb$(\gamma,\,n)$ photoneutron reactions populating the residual nucleus $^{207}$Pb in its (a) ground state, (b) first and (c) second excited states and (d) in any excited state higher than the second. These were obtained by multiplying the branching ratio corresponding to the respective state with the total photoneutron cross section.

The present estimations for the partial cross section show reasonably good agreement with the results of Bell~$et$~$al.$~\cite{bell_1982} (full red dots) obtained at the Illinois bremmstrahlung monochromator using time of flight measurements at seven angles relative to the photon beam. The present $\sigma(\gamma,\,n_0)$ and $\sigma(\gamma,\,n_1)$ cross sections for leaving the residual $^{207}$Pb in its ground state and in its first excited state, respectively, reproduce the narrow peak at 10~MeV excitation energy and the structure at $\sim$10.5~MeV. However, the present $\sigma(\gamma,\,n_2)$ cross section for populating $^{207}$Pb in its second excited state doesn't reproduce the strong structure observed by Bell at 10.5~--~11~MeV. The present $\sigma(\gamma,\,n_>)$ for leaving $^{207}$Pb in an excited state higher then the second is in agreement with the increasing behavior of the Bell data, however with systematically higher absolute values. This follows from the low energy neutron detection threshold of $\sim$600~keV from the Bell experiment, in which low energy neutrons emitted to high excited states were not recorded. 

The present $\sigma(\gamma,\,n_0)$ estimations are further compared with the high resolution measurements of Sherman \emph{et al.}~\cite{sherman_1975}, which were performed using bremsstrahlung beams and the time of flight technique with a single neutron detector placed at 90$^\circ$ to the photon beam. The approximate angle integrated cross sections for populating the ground state of $^{207}$Pb shown by the orange histogram in Fig.~\ref{fig_cs_partiale}(a) were obtained by assuming $W(\theta)=2-P_2(\cos\theta)$ as the neutron angular distribution. Overall good agreement is found between the present $\sigma(\gamma,\,n_0)$ estimations and the data of Sherman with the exception of the 10.5~--~11.5~MeV excitation energy region, where the present results are systematically higher than the ones of Sherman.  

The partial cross sections of Calarco~\cite{calarco_1969} shown by the green crosses in Fig.~\ref{fig_cs_partiale}~(a), (b) and (c) are also approximations. They have been obtained by multiplying by 4$\pi$ the differential cross section measured at 115$^\circ$, in the assumption of isotropic neutron emission. Considering only dipole photoexcitations, s-wave neutron emission is possible from the 1$^-$ states in $^{208}$Pb to the 1/2$^-$ ground state and to the 3/2$^-$ second excited state in $^{207}$Pb. Indeed, there is remarkably good agreement between the present $\sigma(\gamma,\,n_2)$ cross sections and the ones of Calarco. However, the present $\sigma(\gamma,\,n_0)$ cross sections are systematically higher than the ones of Calarco. For the $\sigma(\gamma,\,n_1)$ cross section, the Calarco results confirm the $\sim$11.25~MeV peak position also observed in the present estimations but do not show the structures at 10.5~--~11~MeV present in the Bell measurements and in the present results.   

\section{Statistical model calculations} \label{sec_STAT_calc}

The new experimental data are now compared with statistical model calculations obtained by the EMPIRE \cite{herman_2007_empire} and TALYS \cite{koning_2023_talys} codes. Since the entrance channel plays a fundamental role for an accurate description of the various reaction channels, the fit to experimental photoabsorption cross sections has been tested with several Lorentzian-type closed-forms (SLO, MLO1, SMLO) plus the quasi-deuteron contribution for the E1 $\gamma$-ray strength functions~\cite{capote2009_ripl,plujko_2018}. We found that the SMLO model, properly tuned, reproduces well, though not perfectly, the experimental data. The resulting photon strength function is obtained by the sum of three SMLO-type Lorentzians for which the parameters are given in Table~\ref{tab_gdr}. In addition,  the quasideuteron contribution to the photo-absorption cross section is included with a normalisation factor of 60\% to ensure a proper description of the high-energy tail of the cross section at the highest energies considered here. The resulting SMLO fit is shown in Fig.~\ref{fig_cs_abs} to be rather accurate though some fine structures (e.g. around 20-24~MeV) could not be described.

The specific optical model  potential (OMP)  for $^{206-208}$Pb (RIPL ID - 102)~\cite{capote2009_ripl,Finlay84} is used in the TALYS code to obtain the transmission coefficients for neutron emission. Since the EMPIRE code can only consider a single OMP common to all nuclei in the reaction chain, the Koning-Delaroche general spherical OMP (RIPL ID - 2405) is used in the EMPIRE calculations. The latter was also used for light Pb isotopes within the TALYS code. The level densities at the equilibrium deformation
are described with the enhanced generalized superfluid model \cite{herman_2007_empire} in the EMPIRE code and with the temperature-dependent Hartree-Fock-Bogolyubov (HFB) plus combinatorial model \cite{hilaire_2012} in the TALYS code. In the latter code, the first 40 discrete excited levels coming from experiment \cite{capote2009_ripl} are adopted for all target and residual nuclei. 

In both approaches, the level densities have been adjusted to reproduce the low-lying discrete level scheme and the $s$-wave resonance spacing at the neutron separation energy. In order to reproduce the present experimental results, we found it necessary in the EMPIRE calculations to reduce the so determined $^{207}$Pb level density by lowering the $\tilde{a}$ parameter by 20\%. This increased the $(\gamma,\,1nX)$ cross section tail at the expense of the newly opened $(\gamma,\,2n)$ channel and lowered the $E_{1nX}$ average neutron energies. In the EMPIRE calculations, we also lowered the starting energy for the continuum in $^{207}$Pb from the 4.25~MeV RIPL-3 recommendation to 3 MeV. This improved the description of the $(\gamma,\,n)$ average neutron energies at 10 to 12 MeV excitation energy.

The high photon energies considered in the present experiment also induce an important contribution from the pre-equilibrium cross section. In the TALYS code, it is given by the default two-component exciton model \cite{koning_2023_talys} where only the single-particle state densities of the Pb isotopes have been adjusted to improve the description of the present experimental cross sections. We found that a reduction of about 30\% is needed, except for $^{207}$Pb for which the reduction factor is about 80\%. Such a high reduction factor can be attributed to the strong shell effect in this doubly magic region which is known to strongly affect the level density parameter. 

\begin{table}[b]
\caption{\label{tab_gdr} $^{208}$Pb GDR parameters adopted within the SMLO model for the three Lorentzians. $\sigma$ corresponds to the peak cross section, $E$ to the centroid energy and $\Gamma$ to the full width at half maximum. $i$ is the Lorentzian index.}
\begin{ruledtabular}
\begin{tabular}{lccr}
$i$ & $\sigma_{\rm GDR}^{(i)}$ (mb) & $E_{\rm GDR}^{(i)}$ [MeV] & $\Gamma_{\rm GDR}^{(i)}$ [MeV] \\ \colrule 
1   & 260.65                        & 12.20                     & 3.251                          \\
2   & 526.13                        & 13.93                     & 3.06                           \\
3   & 8.44                          & 25.56                     & 1.97                           \\ 
\end{tabular}
\end{ruledtabular}
\end{table}

The new experimental photoneutron cross sections are compared in Fig.~\ref{fig_cs_gxn} with EMPIRE and TALYS calculations. The $^{208}$Pb$(\gamma,1nX)$ cross section is rather well described by both calculations, though, as for the photoabsorption cross section, some detailed pattern could not be reproduced with the SMLO input photon strength function. Both the EMPIRE and TALYS calculations tend to overestimate the $^{208}$Pb$(\gamma,2nX)$ cross section in the 15 to 20~MeV incident energy region, but a proper description of the energy dependence in the vicinity of 25~MeV is ensured by the inclusion of the third Lorentzian in the SMLO photon strength function (see Table~\ref{tab_gdr}). While the $^{208}$Pb$(\gamma,3nX)$ cross section can be correctly reproduced by both reaction codes, the $(\gamma,4nX)$ channel is underestimated by TALYS but correctly described by EMPIRE.

In a similar way, the newly measured average photoneutron energies are compared in Fig.~\ref{fig_nen_gxn} with EMPIRE and TALYS calculations. Here, we plot the average energies of the exclusive neutron spectra obtained in the EMPIRE and TALYS calculations, which contain the summed contributions of reactions with emission of a given $i$ neutron multiplicity.  The EMPIRE and TALYS model calculations describe qualitatively well the $E_{inX}$ behavior at the lowest incident photon energies. However, the calculations could not reproduce the detailed patterns observed in the energy dependence of $E_{inX}$, e.g. the position of the inflection point at $\sim$11~MeV in the $E_{1nX}$ evolution, suggesting that a better knowledge of the discrete levels in $^{207}$Pb is required to be able to describe the high average neutron energies in the 10 to 12 MeV excitation energy region. Above 25~MeV, the $E_{1nX}$ average neutron energy shows a long decrease, a tendency qualitatively reproduced by the EMPIRE calculations, which however overestimate the $E_{1nX}$ values between 17 and 35~MeV excitation energy and cannot describe the structures present at 16, 20 and 29~MeV. At these high energies, above their corresponding thresholds, EMPIRE also predicts much larger photoneutron energies compared with TALYS, for each of these $E_{inX}$ channels. EMPIRE calculations follow closely the increasing $E_{2nX}$ and $E_{3nX}$ experimental values, but fail to reproduce their average energy drop at high incident photon energies above $S_{4n}$. A decrease in the $E_{2nX}$ is predicted in the EMPIRE calculations at incident energies of~35 MeV, but this follows a continuous increase in average neutron energies up to 6 MeV, well above experimental values. For the $(\gamma,1nX)$ and $(\gamma,2nX)$ reactions, the drop in mean neutron energies predicted by EMPIRE calculations is due to the contribution of charged particle emission channels. This can be seen from the continuous increasing trend of the average energies of neutrons emitted in the reactions without emission of charged particles $(\gamma,\,n)$ and $(\gamma,\,2n)$, represented by thin solid lines in Figs.~\ref{fig_nen_gxn}(a,~b). The sudden increase of $E_{4nX}$ some 4~MeV above the $S_{4n}$ threshold could not be reproduced by the model calculations. 
The difference between both EMPIRE and TALYS predictions stems mainly from the different nuclear level density prescriptions considered and the impact of their associated energy-dependent shell effect. In particular, a smaller shell effect in the level density tends to increase $E_{1nX}$ and decrease $E_{2nX}$ predictions.

The $(\gamma,\,n)$ partial cross sections to the first excited states in $^{207}$Pb are compared with EMPIRE and TALYS calculations in Fig.~\ref{fig_cs_partiale}. As statistical codes, the fine structures observed experimentally cannot be reproduced. However, both TALYS and EMPIRE satisfactorily reproduce the global trend and order of magnitude of the partial cross sections of populating the residual $^{207}$Pb  in its ground and first two excited states. In particular, the TALYS calculations describe well these three partial sections in the region of high incident energies, above 13 MeV, where the statistical neutron emission is dominant. The experimental cross sections $\sigma(\gamma,n_>)$ for populating $^{207}$Pb in any state higher than the second state are relatively well described by both codes.

Finally, based on the newly measured $E1$ photoneutron cross sections, it is possible to estimate the three main moments of the $E1$ distributions, namely 
\begin{itemize}

\item the integrated cross section $\Sigma_{TRK}$ defined in terms of the Thomas-Reiche-Kuhn (TRK) sum rule $\sigma_{TRK}=60 NZ/A$~mb~MeV, {\it i.e.}

\begin{equation}
\Sigma_{TRK}=\frac{1}{\sigma_{TRK}} \times \int_{0}^{\infty} \sigma_{abs}(\omega) d\omega
\end{equation}

\item the centroid energy
\begin{equation}
E_{c}=\frac{\int_{0}^{\infty}\sigma_{abs}(\omega) d\omega}{\int_{0}^{\infty} \sigma_{abs}(\omega) / \omega ~d\omega}
\end{equation}

\item and the polarizability
\begin{equation}
\alpha_D=\frac{\hbar c}{2\pi^2 e^2} \int_{0}^{\infty}\frac{\sigma_{abs}(\omega)}{\omega^{2}} d\omega ~.
\label{eq:pol}
\end{equation}
\end{itemize}
As detailed in Ref.~\cite{goriely_2020}, to estimate the moments, the measured photoabsorption cross sections above the neutron separation energy has been supplemented with the SMLO values below the neutron separation energy and the quasideuteron contribution has been excluded. The results are given in Table~\ref{tab_pol}  where the statistical uncertainties have been simply calculated considering the maximum and minium values of the measured cross section. Table~\ref{tab_pol} also compares the present $E1$ moments with the values obtained in Ref.~\cite{goriely_2020} on the basis of experimental data from Ref.~\cite{tamii_2011}. The resulting moments are seen to be in rather good agreement and confirm the previous mysterious kink found between $^{208}$Pb and $^{209}$Bi polarizabilities in Ref.~\cite{goriely_2020}.

\begin{table}[b]
\caption{\label{tab_pol} $^{208}$Pb experimental integrated cross section $\Sigma_{\rm TRK}$, centroid
energy $E_c$, and polarizability with their estimated
uncertainties (Err) on the basis of the present measurements and comparison with the values obtained in Ref.~\cite{goriely_2020}}
\begin{ruledtabular}
\begin{tabular}{lcc}
 & Present & Ref.\cite{goriely_2020} \\
\colrule
 $\Sigma_{\rm TRK}$ & $1.27 \pm 0.10$ & $1.29 \pm 0.03$ \\
 $E_c$ [MeV] & $14.20 \pm 0.12$ & $14.53 \pm 0.29$ \\
  $\alpha_D$ [fm$^3$/$e^2$] &$20.00 \pm 1.30$ & $19.82 \pm 0.49$  \\ 
 \end{tabular}
\end{ruledtabular}
\end{table}

\section{Summary and conclusions} \label{sec_summary} 

New measurements of photoneutron reactions on $^{208}$Pb have been performed in the GDR energy region using 7.50~MeV~--~38.02~MeV quasi-monochromatic laser Compton scattering $\gamma$-ray beams of the NewSUBARU synchrotron radiation facility. A high-and-flat efficiency moderated neutron detection array of $^3$He counters together with an associated neutron-multiplicity sorting method have been employed.

We obtained the cross sections for the $(\gamma,\,1nX)$, $(\gamma,\,2nX)$, $(\gamma,\,3nX)$ and $(\gamma,\,4nX)$  photoneutron reaction channels. The photoabsorption cross sections, extracted as the sum of the photoneutron cross sections, confirm the Saclay results of Ref.~\cite{veyssiere_1970}. The present experiment extended the area of investigation of the photoabsorption excitation function above the maximum limit of 21 MeV reached in the Saclay experiment. Fine structures have been observed in the $^{208}$Pb$(\gamma,\,n)$ reaction at incident energies lower than 13~MeV and compared with previous high resolution measurements. Average energies of neutrons emitted in each reaction have also been extracted based on the ring ratio data. Low average energies for the $(\gamma,\,1nX)$ and $(\gamma,\,2nX)$ neutrons at excitation energies above $S_{3n}$ indicate a significant contribution of charged particle emission reactions. The experimental photoneutron cross sections and average energies have been satisfactorily reproduced with both EMPIRE and TALYS calculations by slight adjustments of model parameters.

Based on the measured ring ratio values, we have extracted estimations on the total neutron emission spectra in the $^{208}$Pb$(\gamma,\,n)$ reaction at incident energies lower than $S_{2n}$, which confirm a gradual transition from discrete energy neutron emission to a statistical neutron emission towards $S_{2n}$. Present estimations for the partial photoneutron cross sections for populating the $^{207}$Pb residual in its ground and first two excited states reproduce the resonant structures observed in the previous experiments and significantly extend the previously investigated energy range. Until detailed time of flight measurements are available, such systematic estimations on a wide energy range are useful for microscopic descriptions of the GDR structure. 

Finally, the $E1$ moments  extracted from the present data are in rather good agreement with those obtained from the previous measurement of Ref.~\cite{tamii_2011} and confirm the  mysterious kink previously found between $^{208}$Pb and $^{209}$Bi polarizabilities in Ref.~\cite{goriely_2020}.

All the experimental results obtained in the present paper are available in numerical format in the Supplemental Material~\cite{supplemental_material}.

\section{Acknowledgments}
The authors are grateful to H. Ohgaki of the Institute of Advanced Energy, Kyoto University for making a large volume LaBr$_3$:Ce detector available for the experiment. I.G. expresses her gratitude to Prof. M.~Sin of the University of Bucharest, Romania, for her interest in this paper and valuable discussions.  
This work was supported by a grant of the Ministry of Research, Innovation and Digitization, CNCS - UEFISCDI, project number PN-III-P1-1.1-PD-2021-0468, within PNCDI III. 
This work was also supported by the Deutsche Forschungsgemeinschaft (DFG, German Research Foundation) Project IDs 279384907 (SFB 1245) and 499256822 (GRK 2891) and by the State of Hesse under grant "Nuclear Photonics" within the LOEWE program (LOEWE/2/11/519/03/04.001(0008)/62). 
S.G. acknowledges financial support from the Fonds de la Recherche Scientifique (F.R.S.-FNRS) and the Fonds Wetenschappelijk Onderzoek - Vlaanderen (FWO) under the EOS Project nr O000422F. 
F.L.B.G., T.E., V.W.I. and L.G.P. acknowledges funding from the Research Council of Norway through its grant to the Norwegian Nuclear Research Centre (Project No. 341985).

\bibliographystyle{apsrev4-1}
\bibliography{newsubarubibfile}

\end{document}